\documentclass[preprint, 1p, times, authoryear]{elsarticle}
\newcommand{\marginparo}[1]{}
\newcommand{\qnote}[1]{}
\newcommand{\subsectiono}[1]{}
\newcommand{\tableofcontentso}{}
\usepackage{setspace}
\newcommand{\thankso}[1]{}
\newcommand{\footnoteso}[1]{}
\newcommand{\markbotho}[2]{}
\newcommand{\maketitleo}{}

\newenvironment{keywordso}{\begin{keyword}}{\end{keyword}}
\newenvironment{frontmattero}{\begin{frontmatter}}{\end{frontmatter}}
\usepackage[none]{hyphenat}

\usepackage{graphicx}
\usepackage{colortbl}
\usepackage[cmex10]{amsmath}
\usepackage{url}
\usepackage{rotating}

\markbotho{To be submitted}{}

\begin{document}
\begin{frontmattero}
\title{A review of EO image information mining}
\author{Marco Quartulli}
\author{Igor G. Olaizola} 
\address{Vicomtech-IK4, Mikeletegi Pasealekua 57,
Parque Tecnol\'ogico,
20009
Donostia / San Sebasti\'an,
Spain\\
Tel:	+[34] 943 30 92 30,
Fax:	+[34] 943 30 93 93,
E-mail: m.quartulli@vicomtech.org}
\date{\today}
\maketitleo
\thankso{The authors are with Vicomtech-ik4, Donostia, Spain.
Manuscript received \today; revised \today.}

\begin{abstract}
  We analyze the state of the art of content-based
  retrieval in Earth observation image archives
  focusing on 
  complete systems showing promise
  for operational implementation.
  The different paradigms at the basis of the main system
  families are introduced.
  The approaches taken are considered, focusing in particular on the
  phases after primitive feature extraction. 
  The solutions envisaged for the issues related to feature
  simplification and synthesis, indexing, semantic labeling are
  reviewed.
  The methodologies for query specification and execution are
  evaluated.
  Conclusions are drawn on the state of published research in Earth
  observation 
  (EO) mining.
\end{abstract}

\begin{keywordso}
  Remote sensing, Content-Based Image Retrieval
\end{keywordso}

\end{frontmattero}

\tableofcontentso

\section{Introduction}
\label{sec: introduction}
\marginparo{The problem with EO archives}
Earth observation (EO) archive volumes are slowly
approaching the zettabyte scale\footnoteso{
  The data volume for the EOC DIMS Archive in Oberpfaffenhofen is
  projected to about 2 petabytes in 2013 (Christoph Reck, DLR-DFD,
  presentation during ESA EOLib User Requirements workshop, ESRIN
  November 17, 2011)}.
The assets they contain are largely under-exploited: the
majority of records have never been accessed\footnoteso{
  Up to 95\% of records have
  never been accessed according to figures reported in conferences}.
The situation is exacerbated by the growing interest in and
availability of metric and submetric resolution
sensors, due to the ever-expanding data volumes and the extreme
diversity of content in the imaged scenes at these scales. 
Data from missions such
as the ESA Sentinels will be open and free, thus a much larger
audience will want to use them.
Interpreters to manually annotate archived content are expensive and
tend to operate in applicative domains with stable, well-formalized
requirements (e.g. the military) rather than on the open-ended needs
of the remote sensing community at large or of broad efforts like
GEOSS \citep{king2011}.

\marginparo{The solution: EO mining}
Semi-automatic tools to analyze, label, summarize the contents of the
archived image products need to be rendered a standard component of
payload ground segments.

\begin{figure}
  \begin{center}
    \includegraphics*[width=.7\textwidth]{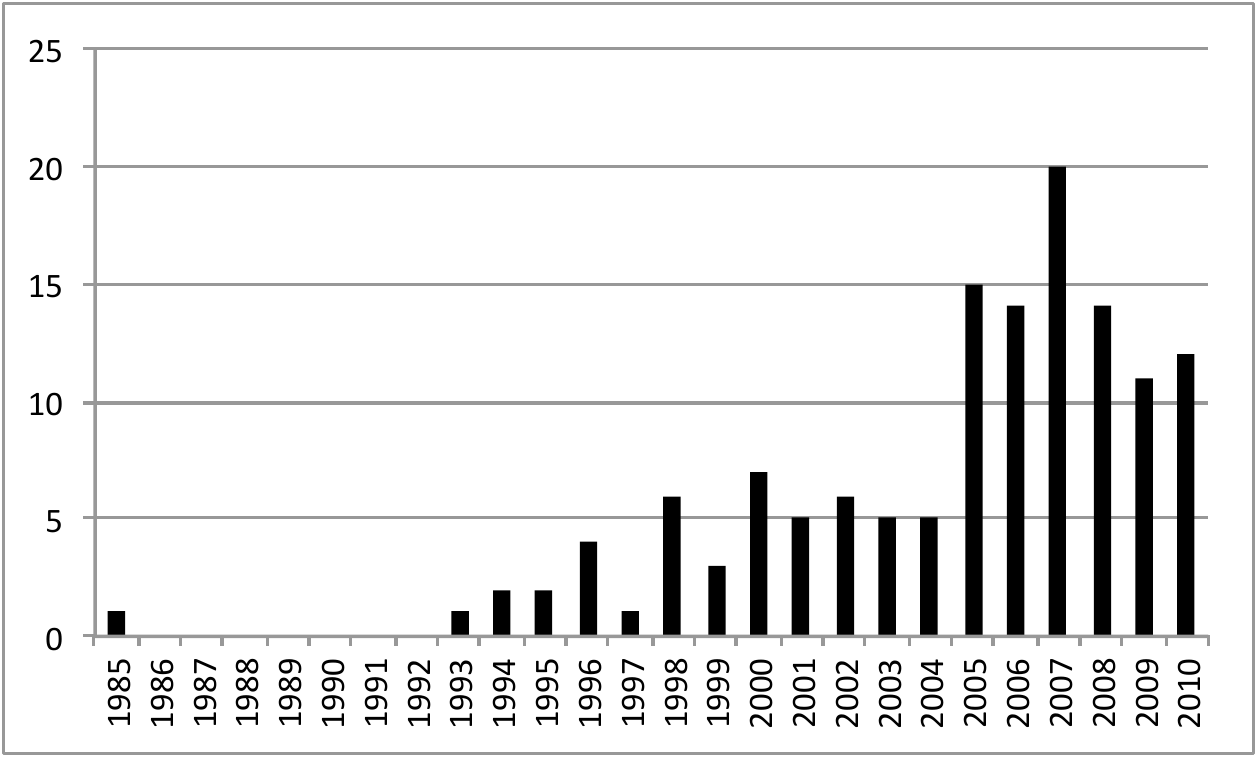}
  \end{center}
  \caption{Papers per year in references database. 
    After a seminal period (up to
    1995), the late 1990s have seen a significant growth of the EO
    mining domain in
    research terms. Years 2000-2005 have focused on a number of
    implementation efforts, with an explosion in the last five years of the
    last decade.}
  \label{fig: cited_per_year}
\end{figure}

\marginparo{Journals and conferences}
Development is currently very active in this domain.
Workshops like the ESA-EUSC-JRC IIM Workshop (\url{http://rssportal.esa.int/tiki-index.php?page=2011_ESA-EUSC-JRC}) are 
organized to discuss about the state of the art in Earth observation
information retrieval.
Special issues of Journals such as the IEEE Transactions on Geoscience
and Remote Sensing have been dedicated to the topic
\citep{datcu2007, datcu_intro_2010}.
Advanced, specialized problems related e.g. to object--based mining
methodologies have been the subject of a number of contributions on
publications such as the ISPRS Journal of Photogrammetry \& Remote Sensing.
The present work analyzes more than one hundred references 
with global geographical representation
and whose temporal
distribution is represented in the plot in figure
\ref{fig: cited_per_year}.

\marginparo{Systems: Region 1-6 --- North America}
In the US, both research oriented systems and operational tools are
under active development.
The first category includes tools like the I$^3$KR at the
Mississippi State University \citep{king2007} and VIS-STAMP at the
University of South Carolina  \citep{guo2003}.
The second one comprises VisiMine \citep{tusk2003}, Insightful
GeoBrowse \citep{marchisio1998}, Earth Perspective's
GeoIRIS \citep{shyu2006} and the RBIR system at Oak Ridge National
Laboratory \citep{tobin2006}.

\marginparo{Systems: Region 8 --- Africa, Europe, Middle East}
European activity in the domain is rich and diverse.
It is embodied in 
systems such as ESA KIM \citep{schroeder_interactive_2000}
\citep{datcu_knowledge_2002}, KEO \citep{__datcu2003}
IIM-TS \citep{bovolo2007}, installed at ESA, DLR, CNES and 
MEA \citep{mantovani2009}, in the Finnish
prototype \citep{molinier2006} based on PicSOM system developed at
Aalto University and in SemQuery \citep{sheikholeslami2002}, 
in the french PLATO \citep{rital2008}, 
the italian Centro Nazionale Multi-missione/Data
Mining component at ASI \citep{garramone2009} and in the upcoming 
TELEIOS EO virtual observatory
(\url{http://www.earthobservatory.eu/}) and in the ESA/DLR EOLib
connected directly to the Sentinel payload ground segments among others.

\marginparo{Systems: Region 9 --- Latin America}
The Brazilian GeoDMA \citep{korting2008, pahl2008}
is currently constrained to the analysis of limited datasets, yet shows
promise for contexts involving larger data volumes.
A further interesting component is TerraLib 
\citep{camara2008}, an open-source GIS
software library that handles spatio-temporal data
types (events, moving objects, cell spaces, modifiable objects) and
allows spatial, temporal and attribute queries on the
database, providing functions for data conversion, display,
exploratory spatial data analysis and spatial and non-spatial queries.

\marginparo{Systems: Region 10 --- Asia and Pacific}
Activity in the Asia and Pacific region is represented by innovative
prototype systems such as those in \citep{ruan2006} and \citep{li2007}.

\marginparo{Transition to operationality}
In the face of all of these research and development efforts, 
archive owners and operators are attempting a rapid transition to operational
systems for automatic data annotation connected directly to satellite
payload ground segments.
Different options are available among architectures and paradigms for
pre-operational Earth Observation digital asset management systems, data
structures and algorithms.
In this paper, we conduct a critical analysis of the state of the art
in this domain, specifically focusing on validated approaches aiming at
operational exploitation, in the hope of providing a
contribution to this transition.

\marginparo{Structure of the paper}
This work is structured as follows: 
section \ref{sec: review_methodology} describes the review
methodology. In section \ref{sec: system_description}, a basic
abstract description of an EO mining system is used to derive a
decomposition of its functionality into modules. The next sections
analyze in detail the different choices taken with respect to specific
processing steps: section
\ref{sec: query_specification_and_processing} describes query
specification and processing, section
\ref{sec: indexing} the indexing approaches, and section
\ref{sec: ingestion} the data ingestion and autonomous analysis.
Section \ref{sec: discussion} reports a general discussion of the
review results.

\section{Review methodology}
\label{sec: review_methodology}
\marginparo{Reference set pooling}
The assessment of the state of the art in EO data mining
has involved discovering, aggregating, normalizing and analyzing
new material to a pool of bibliographic sources in different iterations.

\marginparo{Reference set sources}
The original sources considered have included the journals and the
proceedings of the conferences and workshops mentioned in 
section~\ref{sec: introduction}. 
Items relevant to the query
`(Earth observation OR remote sensing) AND (data mining OR content
based retrieval)' according to
the arXiv at the Cornell University Library (\url{http://www.arxiv.org/}), 
Google Scholar (\url{http://scholar.google.com/}), 
Mendeley (\url{http://www.mendeley.com/}) 
and Zotero (\url{http://www.zotero.org/}) 
have constituted primary material as well.
Progressive aggregations have included 
material cited by these sources.
A further significant source of material has been constituted by
exploring publications citing the papers in the pool via the above
literature search tools and Google Scholar in particular.

\marginparo{Reference set normalization}
The analysis of the characteristics of the pooled material by objective means
has been an important methodological point of the conducted analysis.

\begin{figure*}
  \centering
  \includegraphics[width=\textwidth]{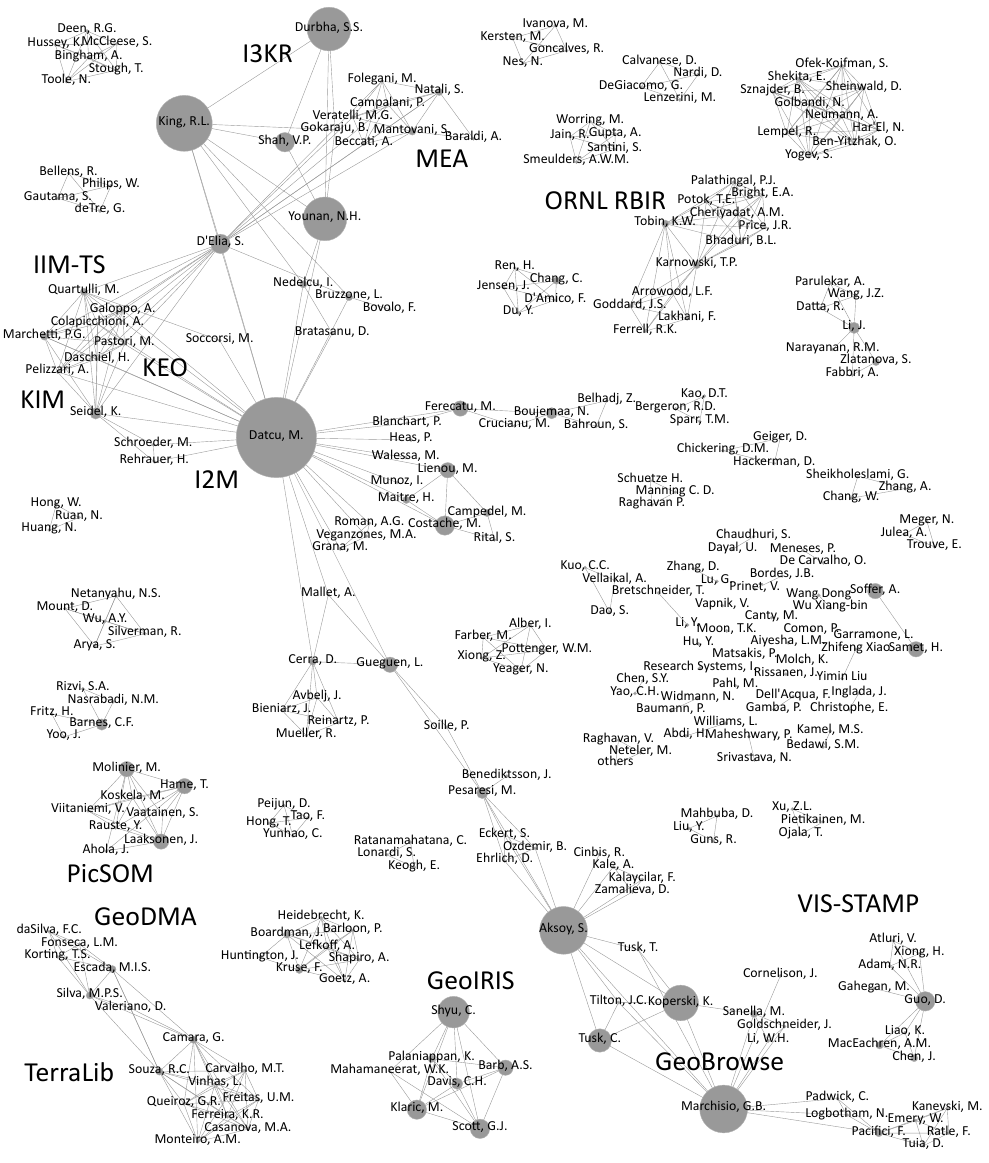}
  \caption{Graph representation of the references database for the
    present review paper. Nodes correspond to single authors, with 
    sizes proportional to number of considered authored publications.
    Edges represent co-authoring relationships, with weights
    proportional to number of joint publications considered.
    The immediately apparent clustering obtained by a Yifan-Hu layout
    optimization procedure \citep{Hu_2005} corresponds to the main EO
    mining system families represented in the referenced literature.
    Framework names in larger letters have therefore been added for
    readability.
  }
  \label{fig: authors}
\end{figure*}

\marginparo{Reference set analysis}
Inspired by \citet{guns2011q} and ~\citet{giuliani2010assessing},
the metadata describing the considered references have been analyzed
by developing a toolset to
build a graph representation of the co-authorship relations among
their authors. The representation obtained by this analysis process
has been manipulated in the open source Gephi~\citep{ICWSM09154}
system (\url{https://gephi.org/})
to obtain the graph map depicted in figure~\ref{fig: authors}.
In it, nodes correspond to single authors. Their sizes are proportional to
the number of authored publications in the references pool, 
while edges, representing co-authoring relationships, 
are weighted proportionally to the considered joint publication number.
The graph representation has then been subjected to a Yifan-Hu layout
optimization procedure \citep{Hu_2005}.

\marginparo{Analysis results}
The immediately apparent clustering resulting in the figure
actually corresponds to the main EO system families represented in the
considered literature:
framework names in larger bold letters have been added to the diagram 
for readability.
The characteristics of the main families of systems in operation 
are summarized together with a description of
their architectures into essential subsystems
in tables~\ref{tab: references_for_systems} and 
\ref{tab: eo_mining_system_decomposition}.
\begin{table}
\centering
\caption{References for the systems in table~\ref{tab: eo_mining_system_decomposition}}
\label{tab: references_for_systems}
\begin{tabular}{lll}
  \multicolumn{1}{ l }{System}&Original Implementation&References\\ \hline
  \multicolumn{1}{ l }{GeoBrowse}&Insightful Corp.&\cite{marchisio1998}\\ 
  \multicolumn{1}{ l }{GeoIRIS}&University of Missouri&\cite{shyu2007}\\
  \multicolumn{1}{ l }{I$^3$KR}&Mississippi State University& \cite{durbha2005c}\\
  \multicolumn{1}{ l }{KEO}& DLR and ACS & \cite{__datcu2003}\\
  \multicolumn{1}{ l }{RBIR}&ONRL & \cite{tobin2006}\\
  \multicolumn{1}{ l }{PicSOM}& Aalto University \& VTT & \cite{molinier2007}\\
  \multicolumn{1}{ l }{MEA} & MEEO & \cite{natali2011}\\
 \hline
\end{tabular}
\end{table}
\thispagestyle{empty}
\begin{sidewaystable*}
\caption{Summary table of main selected EO mining systems with
  capabilities and essential algorithmic components.}
\label{tab: eo_mining_system_decomposition}
\footnotesize
\vspace{1cm}
{\setlength{\extrarowheight}{1.0pt}

\begin{tabular}{|l|l|l|l|l|l|l|l|}
 
\multicolumn{1}{ l }{System}& 
\multicolumn{1}{ l }{GeoBrowse}& 
\multicolumn{1}{ l }{GeoIRIS}&
\multicolumn{1}{ l }{I$^3$KR}&
\multicolumn{1}{ l }{IIM/KIM/KEO/IIM-TS}&
\multicolumn{1}{ l }{RBIR}&
\multicolumn{1}{ l }{PicSOM}& 
\multicolumn{1}{ l }{MEA} 
\\
 



\hline 

\multicolumn{1}{ l }{Sensor}&
\multicolumn{1}{ l }{Multi-spectr}& 
\multicolumn{1}{ l }{Multi-spectr}&
\multicolumn{1}{ l }{Multi-spectr}&
\multicolumn{1}{ l }{Multi-spectr}& 
\multicolumn{1}{ l }{Multi-spectr}&
\multicolumn{1}{ l }{Multi-spectr}& 
\multicolumn{1}{ l }{Multi-spectr}
\\
 
\multicolumn{1}{ l }{classes}&
\multicolumn{1}{ l }{}& 
\multicolumn{1}{ l }{}&
\multicolumn{1}{ l }{}&
\multicolumn{1}{ l }{SAR, Time Series}&
\multicolumn{1}{ l }{}&
\multicolumn{1}{ l }{SAR, Time Series}& 
\multicolumn{1}{ l }{}
\\

\multicolumn{1}{ l }{Res. scale (m)}&
\multicolumn{1}{ l }{10}&
\multicolumn{1}{ l }{1}&
\multicolumn{1}{ l }{10}&
\multicolumn{1}{ l }{1-10}&
\multicolumn{1}{ l }{1}&
\multicolumn{1}{ l }{1-10}&
\multicolumn{1}{ l }{10}
\\

\hline 

\multicolumn{1}{ l }{User interface}&
\multicolumn{1}{ l }{Desktop app}& 
\multicolumn{1}{ l }{WebUI + Google}&
\multicolumn{1}{ l }{WebUI}&
\multicolumn{1}{ l }{Java-based desktop}& 
\multicolumn{1}{ l }{Desktop app}&
\multicolumn{1}{ l }{Unreported}& 
\multicolumn{1}{ l }{Web UI}
\\
 
\multicolumn{1}{ l }{}&
\multicolumn{1}{ l }{}&
\multicolumn{1}{ l }{Earth invocation}& 
\multicolumn{1}{ l }{}& 
\multicolumn{1}{ l }{Graphical UI}& 
\multicolumn{1}{ l }{}&
\multicolumn{1}{ l }{}& 
\multicolumn{1}{ l }{}\\
%

%
 
\multicolumn{1}{ l }{Search}&
\multicolumn{1}{ l }{Interactive}& 
\multicolumn{1}{ l }{Example patch;}&
\multicolumn{1}{ l }{Semantics;}&
\multicolumn{1}{ l }{+/- example pts;}& 
\multicolumn{1}{ l }{Example}&
\multicolumn{1}{ l }{Example}& 
\multicolumn{1}{ l }{Example}
\\
 
\multicolumn{1}{ l }{paradigm}&
\multicolumn{1}{ l }{data}& 
\multicolumn{1}{ l }{extracted obj. attribs;}&
\multicolumn{1}{ l }{metadata;}&
\multicolumn{1}{ l }{text semantic label}& 
\multicolumn{1}{ l }{patches}&
\multicolumn{1}{ l }{patches}& 
\multicolumn{1}{ l }{area}
\\
 
\multicolumn{1}{ l }{}&
\multicolumn{1}{ l }{exploration}&
\multicolumn{1}{ l }{multi-object; semantic}& 
\multicolumn{1}{ l }{examples}&
\multicolumn{1}{ l }{}& 
\multicolumn{1}{ l }{}&
\multicolumn{1}{ l }{}& 
\multicolumn{1}{ l }{}
\\

\hline 

\multicolumn{1}{ l }{Spectral}&
\multicolumn{1}{ l }{Yes}& 
\multicolumn{1}{ l }{Yes}&
\multicolumn{1}{ l }{Yes}&
\multicolumn{1}{ l }{256 classes}& 
\multicolumn{1}{ l }{Yes}&
\multicolumn{1}{ l }{RGB average}& 
\multicolumn{1}{ l }{}
\\
 
\multicolumn{1}{ l }{Texture}&
\multicolumn{1}{ l }{Yes}& 
\multicolumn{1}{ l }{Haralick}&
\multicolumn{1}{ l }{Wavelets}&
\multicolumn{1}{ l }{Gauss--Markov Fields}&
\multicolumn{1}{ l }{Limited}& 
\multicolumn{1}{ l }{Gabor}& 
\multicolumn{1}{ l }{}
\\
 
\multicolumn{1}{ l }{Shape}&
\multicolumn{1}{ l }{Yes}& 
\multicolumn{1}{ l }{}& 
\multicolumn{1}{ l }{Yes}&
\multicolumn{1}{ l }{Zernike on shapes}& 
\multicolumn{1}{ l }{Edge histograms}&
\multicolumn{1}{ l }{Edge histograms}& 
\multicolumn{1}{ l }{}
\\

\multicolumn{1}{ l }{Object}&
\multicolumn{1}{ l }{Yes}& 
\multicolumn{1}{ l }{}& 
\multicolumn{1}{ l }{}&
\multicolumn{1}{ l }{}& 
\multicolumn{1}{ l }{}&
\multicolumn{1}{ l }{}& 
\multicolumn{1}{ l }{}
\\
 
\multicolumn{1}{ l }{Specialized}&
\multicolumn{1}{ l }{}&
\multicolumn{1}{ l }{Anthropogenic}& 
\multicolumn{1}{ l }{}&
\multicolumn{1}{ l }{User pluggable}& 
\multicolumn{1}{ l }{}&
\multicolumn{1}{ l }{NDVI}& 
\multicolumn{1}{ l }{Stratified 56cl.}
\\
 
\multicolumn{1}{ l }{Multi-temp.}&
\multicolumn{1}{ l }{}& 
\multicolumn{1}{ l }{Seasonal trends}& 
\multicolumn{1}{ l }{}&
\multicolumn{1}{ l }{}& 
\multicolumn{1}{ l }{}&
\multicolumn{1}{ l }{}& 
\multicolumn{1}{ l }{}
\\

\multicolumn{1}{ l }{}&
\multicolumn{1}{ l }{}& 
\multicolumn{1}{ l }{trajectory}& 
\multicolumn{1}{ l }{}&
\multicolumn{1}{ l }{}& 
\multicolumn{1}{ l }{}&
\multicolumn{1}{ l }{}& 
\multicolumn{1}{ l }{}
\\
 
\multicolumn{1}{ l }{Multi-object}&
\multicolumn{1}{ l }{Yes}&
\multicolumn{1}{ l }{Yes}& 
\multicolumn{1}{ l }{}&
\multicolumn{1}{ l }{Pluggable extractors}& 
\multicolumn{1}{ l }{}&
\multicolumn{1}{ l }{}& 
\multicolumn{1}{ l }{}
\\

\hline 

\multicolumn{1}{ l }{Clustering}&
\multicolumn{1}{ l }{S-PLUS}&
\multicolumn{1}{ l }{Entropy-balanced}& 
\multicolumn{1}{ l }{Kernel PCA +}&
\multicolumn{1}{ l }{K-Means}&
\multicolumn{1}{ l }{Geospatial}&
\multicolumn{1}{ l }{}& 
\multicolumn{1}{ l }{}
\\

\multicolumn{1}{ l }{}&
\multicolumn{1}{ l }{statistical environ.}&
\multicolumn{1}{ l }{Trees}&
\multicolumn{1}{ l }{Support Vector Machines}&
\multicolumn{1}{ l }{}& 
\multicolumn{1}{ l }{clustering}&
\multicolumn{1}{ l }{}& 
\multicolumn{1}{ l }{}
\\

\hline 

\multicolumn{1}{ l }{User labelling}&
\multicolumn{1}{ l }{}&
\multicolumn{1}{ l }{Sigmoid functions}&
\multicolumn{1}{ l }{Description Logics}&
\multicolumn{1}{ l }{Bayesian classifier}& 
\multicolumn{1}{ l }{}&
\multicolumn{1}{ l }{Self-organizing}& 
\multicolumn{1}{ l }{Nearest Neighbor}
\\
 
\multicolumn{1}{ l }{Query}&
\multicolumn{1}{ l }{Informix}& 
\multicolumn{1}{ l }{Entropy Balanced}&
\multicolumn{1}{ l }{}& 
\multicolumn{1}{ l }{Sequential DB scan}& 
\multicolumn{1}{ l }{K-dimens. tree}&
\multicolumn{1}{ l }{map baed on}& 
\multicolumn{1}{ l }{DB index}
\\
 
\multicolumn{1}{ l }{}&
\multicolumn{1}{ l }{data blade}&
\multicolumn{1}{ l }{Tree scan}&
\multicolumn{1}{ l }{}&
\multicolumn{1}{ l }{with optimized comp.}& 
\multicolumn{1}{ l }{}&
\multicolumn{1}{ l }{neural network}& 
\multicolumn{1}{ l }{}
\\
 
\multicolumn{1}{ l }{}&
\multicolumn{1}{ l }{}&
\multicolumn{1}{ l }{}& 
\multicolumn{1}{ l }{}&
\multicolumn{1}{ l }{of probability index}& 
\multicolumn{1}{ l }{}&
\multicolumn{1}{ l }{}& 
\multicolumn{1}{ l }{}
\\
\end{tabular}
}
\end{sidewaystable*}
\thispagestyle{empty}

\section{System description and decomposition}
\label{sec: system_description}
\marginparo{UML domain diagram}
\begin{figure}
  \centering
  \includegraphics[width=.5\textwidth]{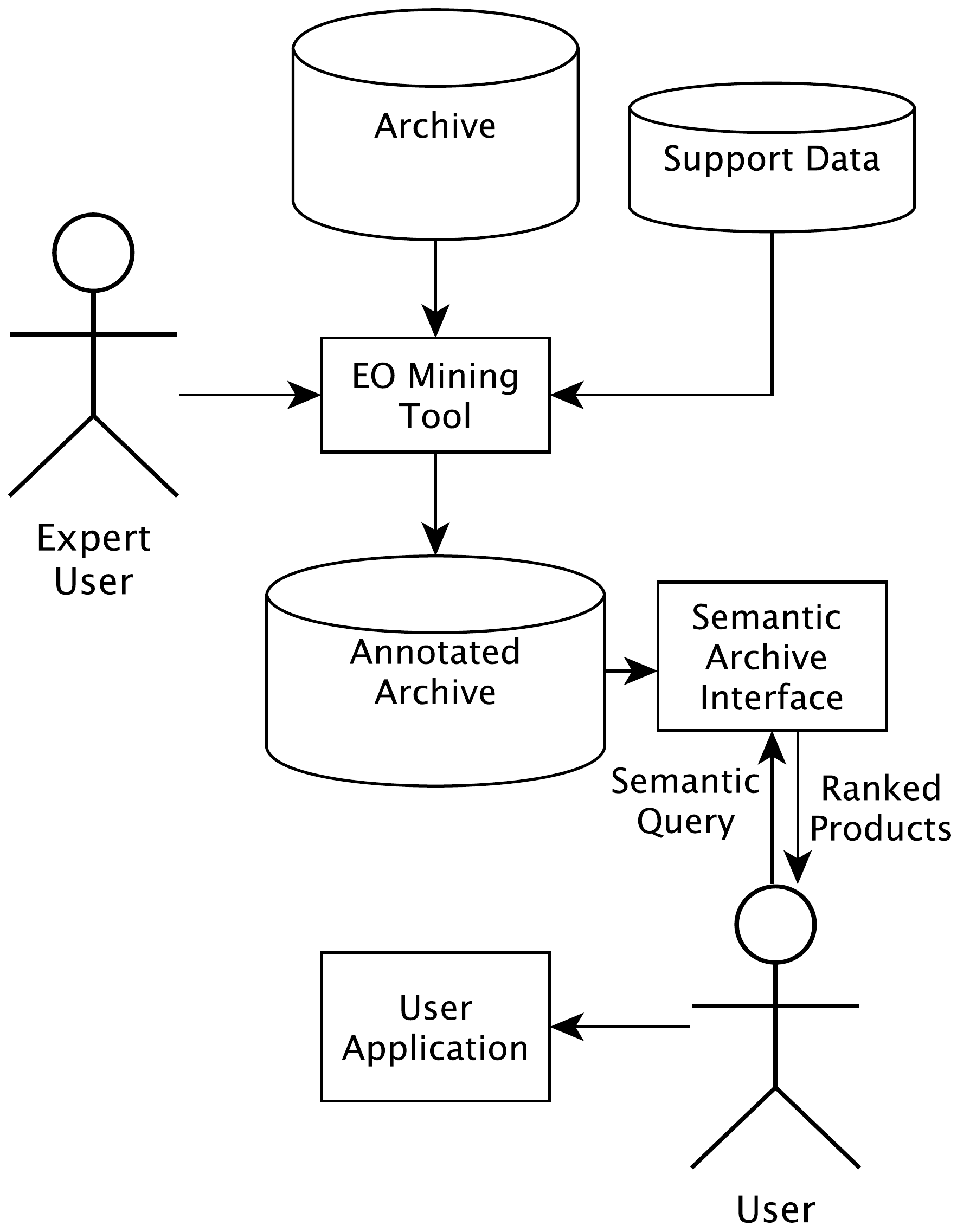}
  \caption{Unified Modeling Language--inspired EO mining context diagram 
    depicting the basic operation of an
    idealized system. Archived images and support data are
    semi-automatically annotated with the help of an expert user. The
    final user queries the annotated version of the archive via a
    semantic interface.}
  \label{fig: eo_mining_context_diagram}
\end{figure}
A few assumptions can be made in order to characterize the 
nature of an EO mining system and of its operating domain
(figure~\ref{fig: eo_mining_context_diagram}).
\begin{enumerate}
\item The system runs on very large (petabyte) scale data base archives.
  Accessible data and metadata are available for the archived products.
\item External support datasets are also often available, from
  pre-existing vector maps to information from additional sensors, and
  might in principle be used e.g. via data fusion methodologies.
\item The main goal of an EO mining system is to help efficiently
  discover, annotate and retrieve specific data products in the
  archive based on
  a semi--automatic characterization of their
  contents / of the contents of the scenes they represent.
\item The overall objective of this capability is to provide support to
  environmental understanding activities such as rapid mapping for
  disaster management, decision support for planning in large scale
  engineering, global climate change mapping.
\item Additional important overall objectives are those related to the
  management and monitoring of the archive and of the whole remote
  sensing system.
\end{enumerate}
\marginparo{Approaches}
To achieve the final goal of EO archive-wide semi-automatic content
discovery, annotation and retrieval, the methodologies envisioned by
the EO mining
systems described in the considered literature references vary from
rapid mapping to archive navigation.
Image processing and analysis, data mining, geographic data
management, object detection and recognition methodologies are relevant,
yet need to be adapted and combined in order to allow characterizing
vast volumes of unknown ---~and at metric scales at least
essentially uncategorizable~--- content.
System structures and algorithmic choices vary significantly as well, from
the considered signal characteristic primitive feature descriptors to
data indexing approaches, from strategies for the optimization of
specific needs for efficient storage access to supervised semantic
label learning and ontology management.
\marginparo{Functionality}
Yet in general we can assume that, in terms of functionality, an EO
mining system
\begin{itemize}
\item at ingestion time, automatically characterizes the input items
  based on a set of characteristics of their data and metadata
  contents that provide a basis  --- the
  primitive feature descriptors --- for all further analysis;
\item semi--automatically annotates the files in an EO archive based on
  their content and metadata. Annotation happens either in terms of
  semantic classes (e.g. `bridge', `house') or in terms of
  application-independent content descriptors (`spectral', `shape') –
  allowing feature synthesis (e.g. mixing spatial/geographic and
  content descriptors) and data fusion (e.g. existing geographical
  maps with EO products);
\item indexes input item sets, building data structures that allow an
  efficient representation of their content for the content discovery,
  annotation and retrieval aims of point 3 above.
\end{itemize}
The above points have consequences that are ubiquitous on the design
and evaluation of EO mining systems.
For instance, it has to be noted that all considerations related to
efficiency in
the remainder of this paper refer to query performance rather than
optimized storage, as is typical in OLAP schemes \citep{chaudhuri1997}.

\subsection{Abstract system model}
\marginparo{The IR problem}
According to \citet{manning2009}, the Information Retrieval (IR)
problem can be stated as the maximization of a certain
utility that ranks visual documents for a user in a specific
context. 
Describing (long term)
user interests about visual information is crucial to the performance of
IR systems. 
To that end, 
relevance can
be expressed explicitly via a rating scale. 
For example, the
binary scale $\{-1, +1\}$ can be used to express ``dislike''/``like''
or ``rejection''/``acceptance'' preferences.
Alternatively, a ``five stars'' scale such as the one
used by Amazon (\url{http://www.amazon.com/}) allows the users to give
more detailed degrees of appreciation.
A Bayesian probability scale $[0, 1]$ can be used to express degrees
of belief in the statement ``data archive item X is interesting to
user Y in the operative and spatio--temporal context
Z''~\citep{jaynes_probability_2003}.
Again in Bayesian terms, semantic analysis and annotation in the
context of EO data mining can therefore be
understood as the procedure of finding the semantic scene elements /
the labels that best describe a given dataset in a given operative
scenario.

\marginparo{IR and Scene Understanding}
We note that this way of expressing the characteristic functionality
of an EO mining tool relates its operation to that of a scene
understanding tool performing inference and estimation 
on the inversion of a direct data acquisition 
model.
When models for the scene 
and the data 
become so complicated
that the inversion 
becomes intractable, a
common solution is a divide-and-conquer approach in which simpler
intermediate description levels are introduced in the modeling.
The scene $S$ is then assumed to be
related to the data $D$ through more levels of intermediate simpler
models $F_i$ that are logically linked
\begin{align*}
  S \rightarrow F_1 \rightarrow \cdots \rightarrow F_n \rightarrow D
  &\mbox{ direct modelling}\\
  D \rightarrow F_n \rightarrow \cdots \rightarrow F_1 \rightarrow S
  &\mbox{ undestanding and retrieval}
\end{align*}
The learning is in this case performed across levels
to derive a conclusion about the underlying scene. 

\marginparo{Smeulders system model}
The generic model corresponding to the treatment given for
instance in the content-based image retrieval review
paper by \citet{smeulders_contentbased_2000} can be adapted to the
context of EO and represented as
\begin{align}
  D \rightarrow N \rightarrow 
  \left\{
    \begin{aligned} 
      O \rightarrow& F \\ 
      G&
    \end{aligned} 
  \right\}
  \rightarrow I \rightarrow
  C \rightarrow S
\end{align}
where
\begin{itemize}
\item $D$ represents the available data
\item $N$ represents a normalized version of $D$, ready for ingestion
\item $O$ represents some form of partitioning of $N$
\item $F$ describes a set of primitive features for the objects in $O$
\item $G$ is a set of signs directly extracted from the normalized
  data $N$
\item $I$ is a joint indexing of the features $F$ and the signs $G$
\item $C$ represents a generic abstract classification of the data
  represented by the index $I$
\item $S$ is a description of the original scene content in terms of
  semantic classes.
\end{itemize}

The way this model architecture translates into algorithmic components
is determined by the system architecture.

\subsection{Architectural options}
\marginparo{Modular systems}
\begin{figure*}
  \centering
  \includegraphics[width=\textwidth]{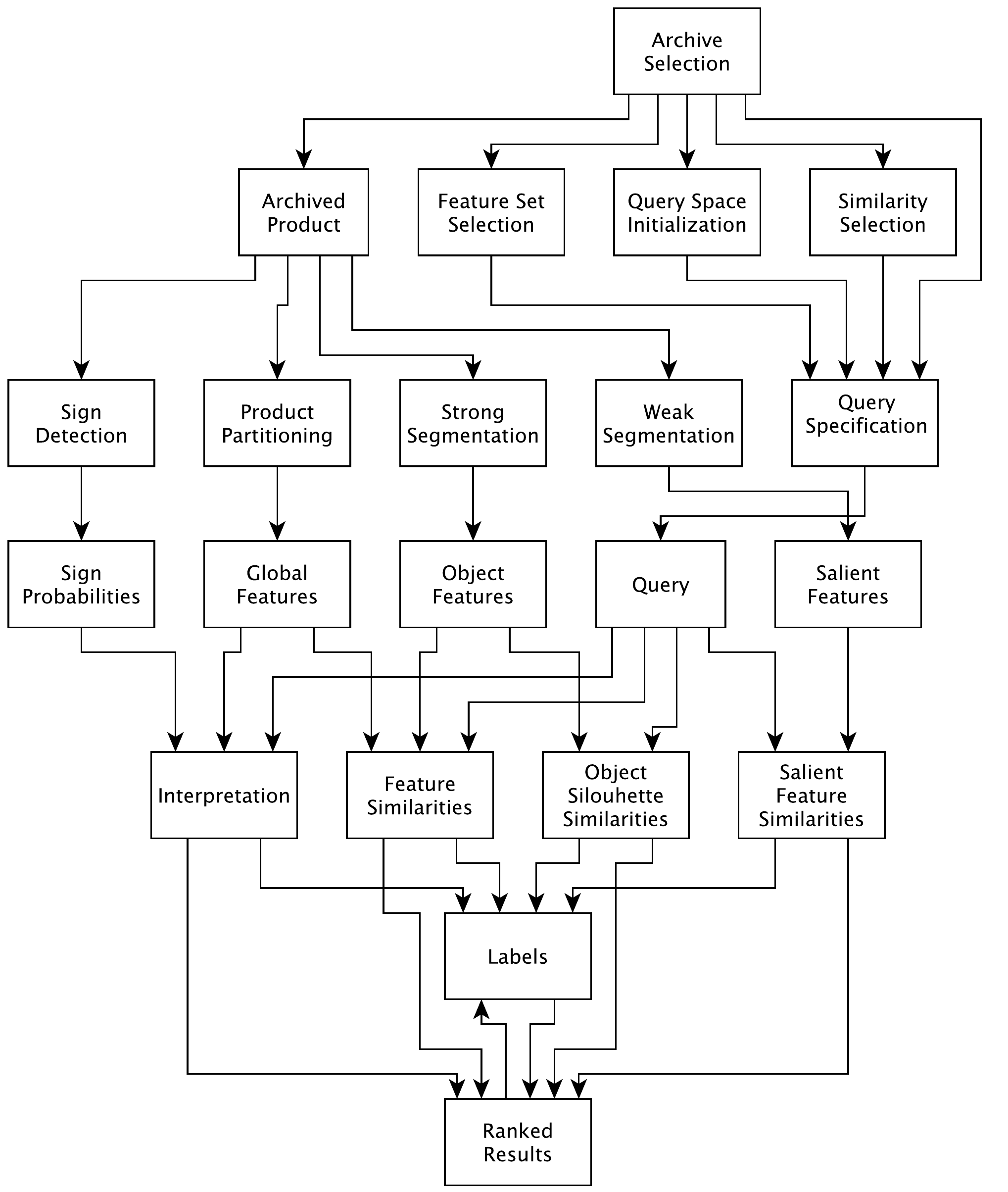}
  \caption{Idealized query process decomposition into processing modules
    and basic operations based on an adaptation of 
    \citet{smeulders_contentbased_2000}.
  }
  \label{fig: system_structure}
\end{figure*}
A general decomposition of a theoretical query process 
is depicted
in figure~\ref{fig: system_structure}.
Archived products can be subject to weak and strong segmentation
processes as well as to a simple partitioning from which regions are
extracted. These regions are then subject to primitive feature
extraction, an
unsupervised data analysis step that generates signatures
in metric spaces 
that express signal characteristics.
Well-known scene elements (``signs'') can be handled by direct
detection and characterization. In the case of EO, these signs
correspond to
characteristic elements of specific interest such as for example
road networks and simple classes of buildings such as silos.
Concurrently, a preliminary selection of an archive collection of
interest and of an appropriate system configuration can lead to the
specification of a query. This query is formulated in such a way as to
be usable for computing similarities in terms of the primitive
descriptors to the analyzed archive datasets. The ranked results can
then be
manipulated by supervised techniques (e.g. including relevance feedback
supervision loops) in order to be used to synthesize labels that
can later be used to enrich the archive contents with semantic
descriptions.

\marginparo{Three stage model}
While the above represents a general view of the operation of an EO
mining tool, 
most of the systems are based on a broad subdivision between an ingestion
component that analyzes the data in an autonomous manner (data
discovery and normalization, primitive feature extraction, indexing),
a learning component that is able to link the primitive feature
information with semantic classes (supervised labeling) and a query
processing system that computes image-to-label and pixel-to-label
distances.

\marginparo{Three stages in actual systems}
The way these main stages are implemented and connected with each
other in actual systems defines their high level architecture.

KEO \citep{__datcu2003} is composed of a number of separate servers with 
SOAP interfaces for much of the
communication both among them and with the user interface.
System web services and interfaces are orchestrated by the
Oracle BPEL Process Manager, to ensure the correct data flow between
modules \citep{munoz2010}.

\marginparo{Communication among stages}
GeoBrowse \citep{marchisio1998}
is based on abstract services and distributed objects. Its operation is
based on the functionality of an object-relational database
management system and of a scientific problem solving
environment, S-PLUS.
Communication between its various components can be
established across platforms and the Internet.

\marginparo{Alternative approaches}
Alternative approaches are also represented.
The RBIR system in \citet{tobin2006}, for instance, breaks down into three
components: (a) a software agent-driven process that can autonomously
search through
distributed image data sources to retrieve new and updated
information, (b) a geo-conformance process to model the data for
temporal currency and structural consistency to maintain a dynamic
data archive, and (c) an image analysis process to describe and index
spatial regions representing various natural and man-made cover types.
Again, the different components are interconnected by web services
with well specified interfaces.

\marginparo{Paper structure}
Moving from architectural descriptions to algorithmic choices, 
we now start analyzing in detail the different choices taken with
regards to specific elements of an EO mining system.
With respect to \citet{smeulders_contentbased_2000}, we prefer
conducting the review in reverse order, starting with the intended
final user operations, since we hope this to better clarify the
different possible tradeoffs.

\section{Query specification and processing}
\label{sec: query_specification_and_processing}
\subsectiono{Query paradigms}
\marginparo{Approaches}
Different approaches exist regarding how a data mining tool for EO
should operate from the point of view of the final user with the aim
of fulfilling the objectives of retrieving datasets of interest and
characterizing them introduced in section~\ref{sec: introduction}.
In \citet{peijun2005}, five retrieval patterns are proposed for EO including
template-based, attribute-based, metadata-based, semanteme-based and
integrated retrieval.
\begin{enumerate}
\item Retrieval based on metadata (e.g. aquisition time, swath
  localization, sensor type). Although this paradigm is not the
  central subject of this review, it needs to be understood as a
  powerful basis for effective browsing and retrieval,
  e.g. contributing to methodologies such
  as those related to faceted search (see below). 
\item Retrieval based on the explicit specification of query
  attributes. In this approach, the query details
  a series of relevant attributes that can be extracted from
  the content of the data items in the searchable base. Examples might
  include both application--independent (e.g. color intervals) and
  specific descriptors (e.g. Normalized Difference Vegetation Index --
  NDVI values).
\item Retrieval based on a template. A user--provided sketch of the
  geometry of the elements of
  interest or a (possibly multi--polygon) relevance mask on a set of
  existing image items or even full image examples are provided as a query
  description.
  A query analysis subsystem analyzes the templates generating a
  description that can be generalized and compared with corresponding
  descriptions extracted at ingestion from the items in the searchable
  data base.
\item Retrieval based on semantemes, or minimal distinctive units of
  meaning, as opposed to sememes, the lower--level units of meaning
  carried by a morpheme that can be considered at the basis of the
  former query specification approaches.
  According e.g. to~\citet{smeulders_contentbased_2000}, a semantic gap needs
  to be taken into account representing
  the lack of coincidence between the information that one can extract
  from the visual data and the interpretation that the same data have
  for a user in a given situation in linguistic, keyword--based,
  contextual terms.
  The semantic gap has its source in the fact that in content-based
  retrieval is that the user seeks 
  semantic similarity, but the database can only provide similarity by
  data processing. The challenge for image search engines ---~
  especially for those designed to operate on a broad
  domain~--- is to tailor the engine to the narrow domain the user has in
  mind via specification, examples, and interaction, 
  effectively creating the relationship
  between semantic information and quantitative descriptions of EO
  signals.
\item Integrated retrieval approaches, in which the complementary
  strenghts of the above methodologies can be combined in order to
  express complex information needs.
\end{enumerate}

\subsection{Interaction patterns}
In practice, these query specification methodologies map to
elements of the user interface and interaction patterns.
\begin{itemize}
\item{Query by textual semantic label}: the user inputs a textual
  description of the scene contents that are being sought.
\item{Query by similarity}: the user provides visual examples of the elements
  being queried. The examples can be pixel-based (the user interface
  allows single clicks to be provided), region-based (the user can
  draw polygons around areas of interest) or tile-based (the input
  products are cut into image tiles with limited size, each of which
  can be used as an example). A further possibility is
  query-by-sketching, in which the user draws a simple outline of the
  elements of interest.
  The examples can belong to a single class or to multiple different
  classes that are considered contextually. Furthermore, some systems
  allow the user to provide both positive and negative examples for
  each class.
\item{Metadata--based faceted search}: 
  A faceted classification system combines data search and
  browsing by classifying each information
  element along multiple explicit dimensions: the user can specify
  complex queries on the parameter space by applying multiple filters
  to explore the query result collection. For instance, a dataset
  might be reduced to a small set of interesting items by
  interactively exploring the results of filtering it with multiple
  metadata-- (e.g. GeoTIFF files only) and content--based (e.g. only
  products containing urban areas) criteria. This search paradigm
  might be implemented in a tree--structured browsing interface
  showing a synthetic statistical description (e.g. the current
  cardinality) of the currently selected product
  set~\citep{Ben-Yitzhak:2008:BBF:}, while allowing
  the user to further filter it, as in the Amazon ``Browse Books''
  catalogue interface (\url{http://www.amazon.com/}).
\end{itemize}
Actual systems mix the different approaches in order to allow users to
express the semantics of the applicative domains they focus on.

\subsectiono{Actual systems}
KIM/KEO \citep{schroeder_interactive_2000, __datcu2003} pre-analyzes all
input products, yet the user can in a later phase interactively
define new cover classes of interest based on positive and negative
example product regions. The classification is only used to provide
visual feedback in the training phase. The obtained semantic labels
are naturally extended to the whole archive, used for ranking search
results and to generate classification maps on selected
products. Systematic processing for classification map generation on
large dataset is also supported.

PicSOM \citep{molinier2007} implements query by visual examples,
with a relevance feedback loop for the interactive learning of
user--level semantics.

GeoIRIS \citep{shyu2007} supports query by example (possibly limited by
region surrounding geographic and anthropogenic features), object and
multiobject spatial relationship queries as well as semantic
queries to answer information needs such as ``given a query image,
show me a set of
satellite images that depict objects and spatial
relationship that are similar to those in the query 
and that are located within a certain radius of a landmark.''
\citet{klaric2006} and \citet{shyu2006, shyu2007} propose
bridging the semantic gap by a three-step process composed of data
transformation (each continuous feature is partitioned into multiple
discrete ranges that are meaningful for a specific semantic), mining
associations (association rules map feature intervals into semantic
categories) and semantic modeling in which crisp intervals from association
rules are replaced by sigmoid functions \citep{barb2007}.

The CNM-DM component \citep{garramone2009} allows users to query images
based on contents via coverage percentages for a pre-defined set of
thematic maps for every ingested product, in an annotation Data-Base
(DB) that is made easily accessible by faceted search
interfaces \citep{Ben-Yitzhak:2008:BBF:}.

Presented in
\citet{marchisio1998, marchisio1999, marchisio2000}, GeoBrowse
is based on the abstract services and distributed objects
paradigm.  GeoBrowse consists of a
Graphical User Interface, an object-relational database
management system, and a scientific problem solving
environment. Each of these can reside on a separate platform.
The system provides
support for intelligent or ``content based''
queries on large databases of remotely sensed images and incremental
and random access to 3-D volumes of multispectral data from different
sensors without the added overhead of multiple storage. GeoBrowse
provides the user with the ability to determine and test the
limitations of remote sensing parameters and models by providing
alternative views of uncertainties arising from extrinsic factors. The
scientific data mining environment is provided by S-PLUS, a commercial
implementation of the S statistical programming language.
A Graphical User Interface (GUI)
offers database browsing capabilities which complement the
functionality of the information retrieval engine through visual
indices and a movie player unit.

Connecting an EO mining system directly to applicative tools allows
users to deal with archive-wide data analysis for objectives such as
large dynamic rapid mapping and multi-temporal scene and sensor
characterization while at the same time leveraging the pattern
discovery abilities and the large data access capacity of the system.
\marginparo{Standardized internal interfaces}
\citet{durbha2008} propose that image information mining systems are
standardized in terms of OGC specifications and in describing the IIM
framework in an OGC perspective. This would facilitate
interoperability with several existing OGC web services and foster the
clear separation of the business logic layer and presentation layer.
OGC standardization efforts related to
geospatial data servers and processing components in particular
promise making interoperability among system components a 
reality \citep{li2007geomatics}.
KEO implements interfaces to and functionalities providing OGC
compliant web services as both an information source
and as a way to distribute the data after semantic processing.
\qnote{add more}

\subsection{Supervised learning \& semantic modeling}
In order to face the
potentially ambiguous meaning of image structures depending on their
contextual understanding, especially in the case of high resolution
remote sensing,
in semantic
search engines based on a
hierarchical information model of satellite image contents a
supervised learning step is used for semantic modeling, to translate the
provided examples into generalized rules for retrieval.

The learned categories group and memorize the
semantics of image structures, facilitating their recognition in
various contexts. Furthermore, the generation of categories helps learning
from a small training data set (i.e. image examples); thus, the method
is useful for the exploitation of very large data volumes,  optimizing
the human machine communication.

\marginparo{Naive Bayesian classifiers}
In the naive Bayesian approach, the assumption is made
that the input features ---~variables
belonging to the same level~--- are mutually
independent and identically distributed (i.i.d.). 
This assumption is not necessarily
justified, yet, naive Bayes results in a simple approach, with clear
semantics, to
representing, using, and learning probabilistic knowledge. It has
often been shown that naive Bayes rivals, and indeed outperforms, more
sophisticated classifiers on many datasets~\citep{duda_pattern_2000,
  witten2011data}, especially when feature selection / synthesis
strategies are put into place in order to better adapt to the i.i.d.
assumption.
KIM \citep{__datcu2003} exploits Bayesian classifiers and Dirichlet models
for defining classes of interest for users. It manages a taxonomy of
semantic labels defined on the hierarchy of signal-based labels.
KIM uses a Bayesian meta-probabilistic measure to link
images to defined labels, whereas GeoIRIS \citep{shyu2007} employs
sigmoid-shaped functions to the same end.
\citet{aksoy2006} uses Bayesian classifiers to compute the final
classification maps using region level information. To be able to use
Bayesian classifiers, different region-based features such as
statistics and shape features are independently converted to discrete
random variables using the K-Means algorithm for vector quantization.

\marginparo{Latent Dirichlet Allocation}
In \citet{lienou2009, lienou2009b}, basic features like means and
variances are computed on a window around each pixel, in each spectral
band. These descriptors are then vector-quantized using a K-Means
algorithm. Semantic annotation of satellite images using the Latent
Dirichlet Allocation (LDA) model. The model
combines a step of supervised Maximum Likelihood classification of patches of the large
image to be annotated, and the integration of the spatial information
between these patches by considering a partial overlap among them.

\marginparo{LDA}
In \citet{lienou2010}, given a training set of images for each
concept, learning is based on the LDA
model. This hierarchical model represents each item of a collection as
a random mixture of latent topics, where each topic is characterized
by a distribution over words. The LDA-based image representation is
obtained using simple features extracted from image ``words''. The
capability of the LDA model to assign probabilities to unseen images
is then used to classify the patches of the large image into the
semantic concepts, using the maximum-likelihood method.

\marginparo{SVMs}
Support Vector Machines (SVMs) provide the basis for a number of systems.
\citet{mountrakis2011support} review remote sensing implementations of
SVMs and compare them with methods from
maximum likelihood classifiers to neural networks in a wide range of
applications from
coal reserve detection to urban growth monitoring.
 
In \citet{durbha2004}, a learning phase is applied at this stage that
associates
the middle level descriptors to the concepts in the higher-level
ontology by means of an SVM \citep{vapnik1999} method. These
associations are grouped into models
specific to a semantic class and used for querying.

\marginparo{SVMs with model selection}
In \citet{li2007b} land cover information corresponding to spectral
characteristics is identified by supervised classification based on
support vector machines with automatic model selection, while
textural features characterizing spatial information are extracted
using Gabor wavelet coefficients. Within identified land cover
categories, textural features are clustered to acquire search
efficient space in an object-oriented database with associated
images stored in an image database.

\marginparo{SVMs and Bayesian inference}
\citet{costache2007, costache2008} propose Bayesian inference to
learn categories and a Support Vector Machine (SVM) classifier to
assign semantics. The approach is enhanced with learning / unlearning
functions.
\citet{Costache_Lienou_Datcu_2006} present a probablistic
extension of SVMs in a Bayesian framework that allows complementing
them with memory simulation mechanisms for machine learning purposes.

\marginparo{SVMs and relevance feedback}
Support Vector Machines are combined with active relevance feedback
in \citet{ferecatu2005, ferecatu2005b, ferecatu2007}.

\marginparo{Decision trees}
Decision trees are also exploited for structur EO data mining.
\citet{aksoy2009} show how decision tree classifiers can be
learned with alternative (surrogate) decision nodes and result in
models that are capable of dealing with missing data during both
training and classification to handle cases where one or more
measurements do not exist for some locations.
\citet{aksoy2010} compute the degrees of satisfaction of the extracted
spatial relationships between objects that satisfy user-specified
attribute criteria. The objects are ranked according to a combined
measure (e.g., product, sum, and weighted sum) that involves the
confidence of detection, the attribute values, and the spatial
constraints.

\marginparo{Bag of words and MDL}
\citet{bordes2008} use
a bag-of-words representation of textons while taking into account
spatial information. A generative probabilistic modeling of the
distribution of textons is proposed. The parameters of the mixture’s
components are estimated using a Expectation-Maximization
algorithm \citep{moon_expectation-maximization_1996}, and the number of
classes in a database estimated by Minimum Description
Length \citep{rissanen_minimum-description-length_1985}.

\citet{koperski2000, koperski2002} describe a system for
interactive training of models for semantic labeling of land
cover. The models are build based on three levels of features: 1)
pixel level, 2) region level, and 3) scene level features. A Bayesian
algorithm and a decision tree algorithm are developed for interactive
training. The Bayesian algorithm enables training based on pixel
features. The scene level summaries of pixel features are used for
fast retrieval of scenes with high/low content of features and scenes
with low confidence of classification. Graphical tools for the
exploration of decision trees allow insight into the interaction of
features used in the construction of models.

\subsection{Iterative query specification}
A common approach allows users to specify examples for the query and
to evaluate the effectiveness of the currently defined one in
successive iterations.

\subsectiono{Interaction: relevance feedback and active learning}
\marginparo{Relevance feedback}
Relevance feedback techniques are used in tools like
KIM~\citep{__datcu2003} and PicSOM~\cite{molinier2007} to improve
retrieval efficiency.
\citet{blanchart_2011} take into account
semi-supervised methods by
mixing an auto-annotation component with a category search engine
which combines generic image class search and object detection.
The proposed concept relies thus on complementary elements: 
an auto-annotation component, a
generic category search engine and an object
detection tool.

\citet{alber2001} shows that a triangle inequality search technique
applied to a relevance feedback retrieval algorithm can significantly
speed up the search for and retrieval of physical events of interest
in large remote-sensing databases. An improvement in retrieval speed
is illustrated using hurricane queries applied to the multispectral
GOES weather satellite database.

\citet{costache2006} present a categorisation-based relevance
feedback search engine for EO images repositories. The developed
method is based on SVMs: the process starts with a query phase in which
the user is selecting among a randomly machine-generated image
sequence one which best describes his interest. The selected image is
then used by the system for retrieval purposes in the following way:
based on a measure of similarity (e.g. Euclidean distance) the system
performs a ranking of the images and returns the top and bottom images.
The returned images are then labelled by the user as
relevant/irrelevant and are used to train an SVM classifier. 
The systems performs again a ranking but this time, based on
the absolute value of the distance function, in ascending order. The
top images with the smallest absolute value of distance function 
are selected to be shown to the user at the next step. They are the
most ambiguous images as they have a very small decision function
value. In this way one step after another the separation surface
between the relevant and irrelevant images is better traced.

\marginparo{Graded relevance feedback}
Non--binary graded relevance feedback is considered by a few authors.
Most of existing methods for annotating semantic meaning to geospatial
images are trained using binary feedback from users. Such approaches
may lead to suboptimal models especially due to the fact that semantic
relevance of images is rarely a binary problem. \citet{barb2010},
\citet{barb2010b} present an algorithm to link low-level image features
with high-level visual semantics using graded relevance feedback from
image analysts. This linkage is done using flexible fuzzy possibility
functions that mathematically model the existence of visual semantics
in new images added to the database.

\marginparo{Active learning}
A number of papers~\citep{ferecatu2007, tuia2009active} focus
instead on active learning for improving remote sensing content based
retrieval: the learning algorithm is able to interactively query the
user to obtain the desired outputs at new data points.

The same approach is used by \citet{rital2008}.
They propose to adapt a classical multimedia CBIR
approach to EO images: they are cut into small images from which a
vectorial signature is extracted. Then an active learning-based
retrieval algorithm is applied in order to profit by the human
expertise. The result of each query can be stored in a memory using
both keywords and classifier model. The system is
evaluated using a small labeled database corresponding to a typical
land cover classification task.

\marginparo{Stepwise retrieval}
In \citet{li2007} a stepwise retrieval scheme is adopted to balance
effectiveness and efficiency. The semantics are used to retrieve a set
of candidate images that are related to the estimated concepts of the
user. Then, the similarity between the query and the candidate images
found in the first stage is measured by integrated region matching.

\subsection{Visualization and visual analytics}
\marginparo{Visualization needs}
The topic of visualization represents an issue with respect to
the main data processing in EO mining whose importance cannot be
overstated:
properly addressing the human visual system of both expert and basic
users in the diagram in figure~\ref{fig: eo_mining_context_diagram}
requires detailed consideration.

\marginparo{Visualization goals}
The goal of visual enhancement methods ---~either by image processing
techniques or by selecting specific spectral bands~--- is to maximize
the response in the human visual system and increase the saliency of
the object / area of interest.

\marginparo{Visualizing data cubes}
\citet{bratasanu2012}
propose a spectral band discovery methodology for advancing
multispectral satellite image visual analysis. The paper describes an
interactive technique to discover the optimum combination of three
spectral features of a multispectral satellite image that enhances
visualization of target classes / objects.

\marginparo{Visualizing sets in abstract spaces}
The inclusion of a Self--Organizing Map (SOM) is one of the central
elements of PicSOM~\cite{molinier2007, molinier2006} 
and of its user interface and key algorithmics.
In it, after a SOM depicting and structuring the data
is trained, the user can visually query the database and the system
automatically finds images similar to those selected.

\marginparo{Visualizing geospatial and multi-temporal objects}
\citet{chen2008} develop a visual analytics approach that
leverages human expertise with visual, computational, and cartographic
methods to support the application of visual analytics to relatively
large spatio-temporal, multivariate data sets. It develops and applies
a variety of methods for data clustering, pattern searching,
information visualization, and synthesis, focusing on combining both human and
machine-based approaches to discover hidden information.

\marginparo{Visual exploration}
Although not directly interested with EO data but rather with
geographic data, \citet{guo2003} describes a human-centered exploration
environment, which incorporates a coordinated suite of computational
and visualization methods to explore high-dimensional data for
uncovering patterns in multivariate spaces. It includes
interactive feature selection and hierarchical clustering methods as
well as a suite of
coordinated visualization and computational components centered around
the above two methods to facilitate a human-led statistical data exploration.
\citet{guo2006}, \citet{guo2006b} integrate computational, visual, and
cartographic methods to develop a geovisual analytic approach for
exploring and understanding spatio-temporal and multivariate
patterns. The developed methodology and tools can help analysts
investigate complex patterns across multivariate, spatial, and
temporal dimensions via clustering, sorting, and
visualization. 
Specifically, the approach involves a SOM, coordinate plots,
several forms of reorderable matrices, a geographic multiple
display, and a 2-dimensional cartographic color design method. The
coupling among these methods leverages their independent strengths and
facilitates a visual exploration of patterns that are difficult to
discover otherwise.

\subsection{Query processing} 
\subsectiono{Query processing} 
%
The query execution process takes in input a set of query items and
returns a set of ranked results.
While execution performance is the primary concern, important
differences arise with respect to the query specification process that
subsequently determine the characteristics of all system components
that prepare the data for this step. Query execution of course
drastically depends on the query
formulation strategy.

\marginparo{Faceted search and fixed classes}
Tools such as the CNM-DM component \citep{garramone2009} only output
thematic maps based on pre-defined classes and coverage percentages
that are fed into an attribute DB.

Most other systems allow users to interactively define new elements of
interest instead.
\marginparo{Query by class, query by value and vector quantization codewords}
A first approach is to allow query by class / query by value based on
vector quantization codewords.
\citet{vellaikal1995a} use vector quantization codewords as
the remote sensed image features for content based
retrieval. Different distortions measures are considered to enhance
the performance of the codewords as ``content descriptors'' including
classification accuracy. Both query by class and query by value are
implemented.

\marginparo{Dynamic queries and relevance feedback}
KIM \citep{__datcu2003} uses a relevance feedback mechanism for updating
a Bayesian network that is used to perform data classification and
retrieval tasks.

\marginparo{Performance and decision tree classifiers}
In the GeoBrowse system, decision tree classifiers are used
\citep{aksoy2004} for interactive learning of land cover models and
mining of image archives: they can operate on both numerical
(continuous) and categorical (discrete) data sources, and do not
require any assumptions about neither the distributions nor the
independence of attribute values.

\marginparo{Semantic web inference}
A methodology to execute complex queries by the integration of an
inference engine is considered by \citet{durbha2005} for $I^3KR$. 
The paper
maintains that pursuing the semantic web model for semantic annotation
of remote sensing data archives provides attractive alternatives to
the traditional methods of information integration and retrieval. It
builds upon semantic web technologies and combines them with
pattern recognition and machine learning techniques to develop a
framework for semantics driven retrieval of knowledge from
EO data archives. At the heart of the framework is a
middleware for ontology brokering that provides support for the
development of application level ontologies that represent the
concepts of the information sources. The Ontology Web Language
 is used to build them. Domain-specific ontologies
help to define concepts in a finer granularity. These fine-grained
concepts then allows to determine specific relationships among
features (e.g., shape, texture, color) in images that may be used to
classify those images.
\citet{durbha2005c} employs an unsupervised segmentation algorithm to
extract homogeneous regions and calculate primitive descriptors for
each region based on color, texture, and shape. They perform an
unsupervised classification by means of a kernel principal components
analysis method, which extracts components of features that are
nonlinearly related to the input variables, followed by a support
vector machine classification to generate models for the object
classes. The assignment of concepts in the ontology to the objects is
achieved automatically by the integration of a description
logics-based inference mechanism, which processes the
interrelationships between the properties held in the specific
concepts of the domain ontology.
In this line, \citet{durbha2005b, durbha2007image} add a methodology for
domain specific qualitative spatial reasoning in coastal wetlands.
In \citet{durbha2007image}, a support vector machines-based classification
is applied for generating predictive models.
\citet{durbha2009} propose a set of feature selection and feature
transformations based on a wrapper-based genetic algorithm approach. A
support vector machine classification is applied for generating
predictive models for those land-cover classes that are important in a
coastal disaster event.

\citet{ruan2006} propose a framework based on domain-dependent
ontology to perform semantic retrieval in image archives.
Homogeneous regions in the data products are described by high-level
concepts depicted and organized by a domain specific
ontology. Interactive learning techniques are
employed to associate regions and high-level concepts. These
associations are used to execute queries. Additionally, a
reasoning mechanism integrating an inference engine
enables mining the inter-relationships among domain concepts and their
properties to satisfy user requirements. An ontology is used to
provide a sharable and reusable concept set as infrastructure for high
level extension such as reasoning.

\marginparo{Multi--stage query execution}
\citet{datta2006} proposes a two-stage architecture for automatic
retrieval of satellite image patches. Semantic categorization is done by a
learning approach involving the two-dimensional multi-resolution
hidden Markov model. Items that do not belong to any
trained category are handled using an SVM--based classifier.

\subsubsection{Data management}
\label{sec: indexing}
\subsectiono{Data base management system options}
Effectively exploiting data storage in order to maximize the potential
of EO mining systems is a central issue for pre-operational and
operational systems. Different approaches are considered in the
literature with respect to this point.

\marginparo{RDBMSs}
While a number of systems such as KIM \citep{__datcu2003} exploit
relational DBs essentially as data storage facilities,
\marginparo{Advanced RDBMSs}
\citet{shyu2006} exploit relational queries directly by representing
in them mining association rule intervals.

\marginparo{NoSQL DBs}
NoSQL DBs \citep{xiao2011} promise scalability, network partitioning,
replication and higher performance than traditional relational systems
for remote sensing image databases. 

An approach to spatial On-Line Analytical Processing (OLAP) based on data
cubes is presented in~\cite{rivest2005solap}. The paper specifically
focuses on interactive spatio-temporal exploration of data via
geovisualization/animation.

\marginparo{Object DBs}
ASI's CNM-DM \citep{garramone2009} and prototype systems at ESA such as
the User Services Next Generation tool already provide faceted
content-based search based on object-oriented databases. 
They exploit a local relational Data-Base Management System (DBMS) for
temporary
object serialization, feeding data into an object oriented DB
for attributes that are then made available for faceted
search.

\marginparo{XML DBs}
The RBIR system described by \citet{tobin2006} exploits XML-based DBs
instead for metadata and
content-based searches. It describes the implementation of a binary
decision tree of the image features based on k--dimensional (kd--)
tree methods
\citep{ayra1994}. The accuracy of the system is selectable as a
trade-off between nearest neighbor performance and computational
efficiency. Through this approach, retrieval efficiencies on the order
of five seconds for 100,000 indexed images can be demonstrated
\citep{tobin2002}.

\subsubsection{Index generation and maintenance}
Indexing is meant to improve the speed of data
retrieval operation by building summary
representations of the content of the database in terms that exploit
the inherent organization of the data to allow very fast query
execution procedures:
the goal of indexing is to organize the image data (e.g., filenames,
features, indexing codes, etc.) in the database such that a ranked
list of nearest neighbors can be efficiently retrieved in response
to a query without performing an exhaustive comparison to all the
records in the database.

\marginparo{No index: sequential scan}
In systems such as KIM \citep{__datcu2003} no indices are stored in
the DB. The ranking
phase operates sequentially on all records by dynamically computing
Bayesian estimates of query-to-tile match based on the selected
primitive features.

\marginparo{Balanced trees}
Whereas an exhaustive
nearest-neighbor search of the $n$ vectors (i.e., images) in the
database would imply of $O(n)$ computations, a kd-tree approach
involves a complexity $O(log_2(n))$. 

In GeoIRIS \citep{shyu2007} two indexing mechanisms are used for tile-
based and object-based query methods. Indexing of continuous valued
features is done using the entropy balanced statistical (EBS)
k-dimensional (k-D) trees, and indexing the binary-valued features is
performed with the entropy balanced bitmap (EBB)
tree \citep{scott2011}, which exploits the probabilistic nature of bit
values in automatically derived shape classes. Tile-based
indexing provides access into localized areas of similar
features. Object-based indexing includes both individual objects and
spatial configurations of multiple objects.

\marginparo{Automatic index generation in column and raster DBs}
Column--based and raster DB systems \citep{ivanova2007monetdb}
\citep{Widmann:1998:EEO:646497.695765} attempt at automatic record
optimization and autonomous index generation. Altough the results
largely improve on sequential scan in traditional relational DBMSs, 
more advanced implementations (e.g. based on k--d trees) are needed
\citep{Kao:1998:EPS:646497.695620} for optimal results.

\section{Ingestion}
\label{sec: ingestion}
\subsectiono{Data discovery}
The first necessary step before data analysis is the discovery of the
data and its normalization so that all subsequent phases of the
analysis process can operate effectively.
This functionality corresponds to crawlers in web information
retrieval engines.

\marginparo{KIM rolling archives}
KEO \citep{schroeder_interactive_2000, __datcu2003} is directly
connected to specific DFD and ESA feeds and rolling archives and continuously
updates its contents respectively with MerisRR and TerraSAR-X data.

\marginparo{RSS feeds}
In a parallel concept, \citet{bingham2009} takes the concept of Really
Simple Syndication (RSS) feeds, for delivering regularly changing web
content, and extend
this to represent a stream of data granules and deliver regularly
changing Earth science data content. It envisages filtering content
based on the metadata of a feed in order to identify granules of
interest based on user-defined criteria.

\subsection{Dataset tiling}
\marginparo{Smeulders' classification}
In terms of \citet{smeulders_contentbased_2000}, a few possible ways
exist to decompose images in order to analyze their contents, each one
corresponding to a different class of content descriptors.
\begin{itemize}
\item No decomposition: sign extraction.
\item Resolution reduction: salient features.
\item Generic tiling: generalistic features.
\item Segmentation: region features.
\item Object extraction: object-based descriptors.
\end{itemize}
The effects of sensor resolution are especially important in
determining which of the approaches provides the most interesting
results.

A tiling approach similar to the one in \cite{rital2008} is used in the
PicSOM system~\citep{molinier2007}. Here, query by visual examples,
relevance feedback and ranking of images
rely on the standard multimedia CBIR approach related to cutting satellite
images into small images (imagelets).
Most systems consider instead geographically--inspired tiling grids.

\marginparo{Hierarchical segmentation}
\citet{trias2008using} review several existing colour space
transformations and textural features for the segmentation of
high-resolution multispectral aerial images with a hierarchical
segmentation algorithm. A method to quantitatively evaluate the
quality of a hierarchical image segmentation based on Pareto
optimality is presented, and the
behaviour of the segmentation algorithm for various parameter sets is
explored.

\marginparo{Wavelet-based segmentation}
\citet{guo2009} proposes a semantic-aware two-stage
wavelet-based image segmentation approach, which preserves the
semantics of real-world objects during the segmentation process.
The system is specifically aimed at high resolution remote sensing
image retrieval.

\marginparo{Segmentation by Differential Morphological Decomposition}
\citet{gueguen2011} presents an interactive image information
mining tool handling millions of structures in a scene. The learning
process is incremental, incorporating new training samples at low
computational cost.
Input images are first segmented in multi scale blobs by segmentation
of Differential Morphological Decomposition, and then are
characterized by spectral and shape features. Both cluster and SVM
based classifiers are compared in terms of accuracy, where the
accuracy is the complement of the pixel based probability of
errors. Classifier training is performed incrementally in both cases,
benefiting from a global classification visualization at each step.

\marginparo{Middle level object ontology}
\citet{durbha2004} employs an unsupervised segmentation algorithm to
extract homogeneous regions and calculate primitive descriptors for
each region based on color, texture and shape. The primitive
descriptors are described quantitatively by middle level object
ontology. 

\subsection{Primitive features and signs}
\subsectiono{Sensor model management}
Ingestion operates by
analyzing the different information sources trying to unsupervisedly
characterize them, often in application--independent manners.
EO mining systems operate on primitive signal features extracted from
the original data. Their performance depends crucially on them and on
their adaptation to specific mining objectives. Generalist primitive
feature extractors are used to make application-independent image
mining possible. These can then be complemented by vertical
application-specific feature extractors dedicated to specific targets,
corresponding to ``sign'' descriptors in
\citet{smeulders_contentbased_2000}.

%
\marginparo{Theoretical introduction}
As per \citet{smeulders_contentbased_2000}, all
systems consider first a primitive feature extraction step that
transposes the image data into another spatial data array. 
The different descriptor extraction methods (e.g. color, local
texture, local geometry) may be characterized in general by:
\[ f(x) = g \circ i(x) \]
where $i(x)$ is the image, element of image space $I$, $g$ is an
operator on images, and the resulting image field is given by
$f(x)$. Computational parameters of $g$ may include the size of the
neighborhood around $x$ to compute $f(x)$ or a homogeneity criterion
when the size of the patch to compute $f(x)$ depends on the actual data,
as in \citet{m._soccorsi_space-variant_2006}, for example.
The purpose of this signal processing in image retrieval must be to
enhance aspects in the image data relevant to the query and to reduce
the remaining aspects.

\marginparo{Invariance}
The aspect of dealing with invariance as a tool to deal with
accidental distortions in the information introduced by the sensory
gap is often handled by primitive classification.
Again according to \citet{smeulders_contentbased_2000}, the aim of
invariant descriptions is to identify objects, no matter from how and
where they are observed, at the loss of some of the information
content. If two objects $t_i$ (or two appearances of the same object) are
equivalent under a group of transformations $W$, they are in an
equivalence class
\begin{equation*}
t_1 \overset{W}{\sim} t_2 \Leftrightarrow \; \exists \, w \in W : t_2 = w \circ t_1 \; .
\end{equation*}
A property $f$ of $t$ is invariant under $W$ if and only if $f_t$ remains the
same regardless the unwanted condition expressed by $W$,
\begin{equation*}
t_1 \overset{W}{\sim} t_2 \Rightarrow f_{t_1} = f_{t_2} \; .
\end{equation*}
Again as noted in the paper cited above, the degree of invariance,
that is, the dimensionality of the group $W$, should be tailored to the
recording circumstances. The aim is to select the tightest set of
invariants suited for the expected set of non constant conditions. 

\marginparo{Primitive feature taxonomy}
We classify the primitive extraction into global features, salient
features, object features and signs according to
\citet{smeulders_contentbased_2000}.

\subsubsection{Signs}
The ability of an IR system to effectively incorporate and exploit an external
primitive feature extractor for search in large archives is a central
factor in defining its adaptability and hence its real-world
applicability.

\marginparo{User pluggable descriptors}
The KEO system \citep{__datcu2003} includes tools to allow users specify
primitive feature extractors dedicated to specific target classes.

\marginparo{Commercial libraries}
Environments such as ENVI/IDL \citep{canty2007image, research2003envi}
include large repositories of image and object
analysis modules. In particular, the ENVI EX system provides feature
extraction, classification, orthorectification and change detection
functions. An integration with the ArcGIS platform allows users to
integrate image analysis processes into GIS models via high--level
programming languages such as IDL.

The eCognition image analysis software is introduced
by~\citet{benz2004multi} with the principal strategies of
object-oriented analysis behind it, discussing how the combination
with fuzzy methods allows implementing expert knowledge and describing
a representative example for the proposed workflow from remote sensing
imagery to GIS.

\marginparo{Open--source feature libraries}
Similarly, QGIS (\url{http://www.qgis.org/}), 
gvSIG (\url{http://gvsig.org/}) and
GRASS (\url{http://grass.osgeo.org/})
offer an open--source alternative \citep{neteler2006advances} for
integrated GIS and remote sensing image analysis that
includes statistical analysis tools based on low-- as well as
high--level programming environments such as 
R (\url{http://r-project.org/}).

\marginparo{Orfeo Toolbox}
\citet{inglada_igarss_2009} describes the Orfeo ToolBox (OTB), a
remote sensing image processing library developed by CNES that
includes tools for the operational exploitation of the future
sub-metric optic and radar images (rapid mapping, 3D aspects, change
detection, texture analysis, pattern matching, optical and radar
complementarity).
A number of the algorithms made available in the OTB can be used as sign
extractors in EO mining engines.


\subsection{Object-- and region--oriented descriptors}
Object--oriented methods \citep{blaschke2010object} are making
considerable progress towards a
spatially explicit information extraction workflow, such as is
required for spatial planning as well as for many monitoring
programmes.

The exploitation of available prior knowledge maps for the extraction
of important topographic objects, like buildings and roads, is
investigated in~\cite{baltsavias2004object}. The paper focuses on
aspects of knowledge that can be used for object extraction: types of
knowledge, problems in using existing knowledge, knowledge
representation and management, current and possible use of knowledge,
upgrading and augmenting of knowledge. An overview on commercial
systems regarding automated object extraction and use of a priori
knowledge is given.

\subsectiono{Sign relationships}
\marginparo{Object spatial relationships}
The GeoIRIS system \citep{shyu2007} supports the retrieval of tiles
according to the spatial configuration of the objects they contain. 
In particular \citet{scott2005} includes a method to model spatial
relationships among sets of three or more objects in satellite images
for scene indexing and retrieval by generating discrete spatial
signatures. The method is highly insensitive to scaling, rotation, and
translation of the spatial configuration.
In \citet{aksoy2010}, directional spatial relationships among objects
are considered to enable object detection based on the properties of
individual objects as well as their directional spatial relationships
to other objects based on morphological modeling of
relative-position-based spatial relationships. This information is
incorporated into the Bayesian decision rule as spatial priors for
contextual classification and retrieval. The directional landscapes
considered can be used for image retrieval for geospatial
intelligence.


CBIR using shapes and topology is considered
in~\cite{agouris1999environment} for sketch--based query formulation.
A spatial data management system optimized for such queries and a set
of features describing the geometry of the
sketched query shapes as well as their spatial arragement are presented.

\marginparo{Higher level objects}
In \citet{guo2009},
to better capture semantic features for object discovery, a
hyperclique pattern discovery method is exploited to find in the DB
co-occurrence patterns representing complex objects that consist of
several co-existing individual objects that usually form a unique
semantic concept.
\marginparo{Object to object graph mapping distances}
In \citet{gautama2007}, spatial relations between objects are used to
find a reliable object-to-object mapping. Graph matching is used as a
flexible query mechanism to answer the spatial query.

\subsectiono{Spatial partitioning}
Except for sign extractors, other kinds of data content descriptors
need data partitioning strategies.
Both strong and weak segmentation can be used to this end, resulting
respectively in salient features and in region based descriptors.

\subsectiono{Region features}
Even though correct identification of pixels and regions improvse the
processing time for content extraction, manual interpretation is often
necessary for many applications because two scenes with similar
regions can have very different interpretations if the regions have
different spatial arrangements. Therefore, modeling spatial
information to understand the context is an important and
challenging research problem.
In \citet{tobin2006}, once the features of the image segments have been
extracted, it is possible to use this feature vector as an index for
retrieval. A geospatial clustering procedure is performed using a region
growing technique to connect large contiguous and homogeneous segments
of similar structure and texture characteristics.
\marginparo{Region quantization}
In \citet{li2007}, every image is segmented into disjoint regions using
a simple segmentation algorithm, and the low-level features are
extracted from each region. In a second step, the region descriptors,
i.e., multidimensional vectors, are classified into a finite number of
frequent patterns with similar appearance by using a vector
quantization algorithm which creates a reduced representation of
possible region descriptors. The images in the database are encoded by
so-called codes based on the result of the previous
quantization.

\marginparo{Shape similarity}
Shape similarity is a powerful tool for query processing.
\citet{dellacqua2001} present a ``point diffusion technique''
method for
shape similarity evaluation suitable for application to
similarity-based retrieval from remotely sensed meteorological image
archives, where shapes are hardly defined but are still among the
major features of interest.
\citet{silva2005} combine a region-growing segmentation algorithm on
multi-temporal change images with a structural classifier that
operates on shape descriptors for mining evolutions typically related
to deforestation patterns.
\marginparo{Region polygonal approximation}
In \citet{aiyesha2005}, from 56 full hemisphere infrared scans of the
earth taken by Meteosat satellites, regions are extracted using region
growing. After region extraction, polygonal approximation is applied
to the region shape, and local features of the polygons are hashed to
provide an association space. This space becomes the indexing
structure through which retrieval takes place.
\marginparo{Statistical region summaries}
\citet{aksoy2006} models regions using the statistical summaries of
their spectral and textural properties along with shape features that
are computed from region polygon boundaries. The statistical summary
for a region is computed as the means and standard deviations of
features of the pixels in that region. Multi-dimensional histograms
also provide pixel feature distributions within individual regions.

\marginparo{Region relationships}
Relationships among regions can be considered as well.
The VisiMine system
\citep{tusk2003, koperski2002b, aksoy2002, aksoy2005}
includes automatic methods for the extraction of topological,
distance-based, and relative-position-based relationships between
region pairs where such relationships can be successfully used for
image classification and retrieval in scenarios that cannot be
expressed by traditional pixel- and region-based approaches.
\marginparo{Region spatial relationship histograms}
\citet{kalaycilar2008} describe an image representation using spatial
relationship histograms that extend relational graphs. These histograms are
constructed by classifying the regions in an image, computing the
topological and distance-based spatial relationships between these
regions, and counting the number of times different groups of regions
are observed in the image. A selection algorithm produces compact
representations by identifying the distinguishing region groups that
are frequently found in a particular class of scenes but rarely exist
in others.


\subsection{Spectral and textural descriptors}
Almost all of the considered systems use spectral and textural
descriptors as well as local edges and patterns. 

GeoIRIS \citep{shyu2007} exploits
histograms for panchromatic, grayscale RGB, and
near-infrared data.
\subsectiono{Textural features}
As textural features, KIM \citep{schroeder_interactive_2000, __datcu2003}
considers Gauss--Markov Random Field-based descriptors (usually from a
single specific
channel) for both optical and SAR data. Advanced Bayesian
reconstruction methodologies \citep{walessa2000model}
are used to cope with the specificities
of SAR sensors.

\marginparo{Gabor}
\citet{aksoy2006} extracts spectral and textural features for each
pixel. Gabor texture features are extracted by filtering the
first principal component image with Gabor kernels at different scales
and orientations and using kernels rotated by $n \pi /4$, $n = 0,
\cdots , 3$, at 4 scales resulting in feature vectors of length
16. Finally, each feature component is normalized by linear scaling to
unit variance.
\marginparo{Edge information, local patterns}
\citet{tobin2006} primarily exploit edge information to describe
texture and structure to avoid performance degradation due to
variation in the spectral content. The approach characterizes segment
texture using local binary patterns \citep{pietikainen2000} and
local edge patterns \citep{yao2003}.
\marginparo{Haralick}
The GeoIRIS system described in \citep{shyu2007} exploits texture
measures based on Haralick co-ocurrence matrices.

\marginparo{Wavelets}
\citet{shah2005, shah2007} propose to perform image 
segmentation using color and texture features from the wavelet
coefficients for the region-based retrieval in remote sensing image
archives.
Based on that, \citet{shah2006} exploit Independent Component Analysis
 \citep{comon_independent_1996} for feature selection/synthesis.

Clustering issues include the possibility to consider different
distance measures among n-dimensional multi-spectral vectors
\citep{cerra2011compression}, ranging
from euclidean \citep{keogh2004towards} to spectral angle
\citep{kruse1993spectral, kruse1999fifteen}, from spectral correlation
\citep{decarvalo2000spectral} to information divergence measures
\citep{du2004new}.
KIM \citep{__datcu2003} uses K-Means clustered multi-spectral feature
histograms based on either euclidean or spectral angle distance
measures.

\subsectiono{Specialized features}
A few systems consider specialized descriptors as well.
\marginparo{High resolution data features}
In the GeoIRIS system \citep{shyu2007}, specialized features are
exploited to represent characteristics of linear features such as
roads (linearity, directionality) and the scale of objects such as
buildings (pixel correlation run-length and scale-based descriptors of
object content generated from differential morphological profiles
\citep{pesaresi2001}). Furthermore, a set of features based on
normalized grid descriptors are extracted for each object
\citep{zhang2001}. The histogram of forces approach \citep{matsakis1998}
is used to generate a spatial signature of an object configuration by
extending the pairwise determination of spatial relationships.
GeoIRIS also includes an ``anthropogenic'' primitive
descriptors to try and separate populated from natural areas.
\marginparo{NDVI}
The PicSOM system \citep{molinier2005, molinier2007} exploits as
features color
moments and textures together with NDVI and edge histograms.

\subsectiono{Multi-sensor}
\subsection{Data fusion}
Data fusion is a significant topic in its own right. \citet{molch2010}
for instance examines the representation of semantic categories
integrating Ikonos and Quickbird imagery in the knowledge-based
information mining system KIM. A processing sequence is presented,
which accounts for sensor-related differences along with an
evaluation of the application of IIM technologies in operational
rapid-mapping scenarios. 

The PicSOM system \citep{molinier2007, molinier2006}, includes
features adapted to man-made
structures detection in high-resolution optical satellite
images. Fusion of panchromatic and multispectral information can be
obtained by training several SOMs in parallel (one per feature).
\citet{molinier2008} investigates the potential of PicSOM for
quad-polarised ALOS PALSAR images.
The system was originally developed to operate on QuickBird
MultiSpectral and Panchromatic data.

\subsection{Multi-temporal mining}
\marginparo{SITS}
Satellite Image Temporal Series (SITS) represent a very active objective of EO
mining research: the extension of the change detection problem
---~including
high geometric resolution data
\citep{padwick2011}~--- to long series of semi-periodic data.

\marginparo{Object--based}
In \citet{walter2004object}, a change detection approach based on an
object--based maximum likelihood classification of groups of pixels
corresponding to objects in an existing GIS database used to provide
training areas.

\marginparo{PicSOM}
The PicSOM system~\citep{molinier2007} implements original methods for
both supervised and unsupervised change detection based on the ranking
of discrimination values for imagelet pairs, and distance on the SOM
lattice. The system is specifically evaluated for the analyzing
variations in the presence of buildings, differentiating
true changes with respect to unsubstantial effects (e.g. changes in
illumination).

\marginparo{Heas}
A seminal work is represented by \citet{heas2005modeling}: a
hierarchical Bayesian modeling of SITS information
content enables uerss to link semantic interests to specific
spatio-temporal structures. The hierarchy is composed of two
inference steps: an unsupervised modeling of dynamic clusters
resulting in a graph of trajectories, and an interactive learning
procedure based on graphs which leads to the semantic labeling
of spatio-temporal structures

\marginparo{IIM-TS}
The IIM-TS system \citep{bovolo2007}, instead, integrates probabilistic
multi-temporal descriptors into the KEO system.

\marginparo{Information theoretical approaches}
On a methodological level, information theoretically motivated approaches are
well represented. \citet{cerra2010} and \citet{gueguen2007} present an algorithm
based on the Information Bottleneck principle composed of a parameter
estimation and a model selection. Two approaches are presented. In the first
approach, each image of the SITS is segmented and the obtained regions
are described by textural models. The Information Bottleneck method is
further used to characterize the image segments of the SITS a
spatio-temporal way. In the second method, the geometrical information
is extracted from a temporal adjacency graph of the spatial regions,
and the radiometric and textural information is then extracted through
the Information Bottleneck method. This approach leads to a temporal
characterization of the spatial regions of the SITS.
In the context of compression-based analysis, \citet{gueguen2008}
present a Satellite Image Time Series coding system that produces a
code in two parts: one contains the predictor and the statistical
context tree, and the other one contains the residual error coded in
an optimal way. This enables the creation of an index of
spatio-temporal structures.

\marginparo{MEA}
\citet{julea2006} proposes an approach similar to that of
\citet{mantovani2009, natali2011} and based on the use of sequential
patterns for the
analysis of multi-temporal remote sensing data. Indeed, as sequential
patterns include the temporal dimension, they can be used for
extracting frequent evolutions at the pixel level, i.e. frequent
evolutions that are observed for geographical zones that are
represented by pixels. While \citet{julea2006} deals with METEOSAT and
ERS satellite images, \citet{mantovani2009, natali2011}
Multi-sensor Evolution Analysis (MEA) is able to
deal with  15 years of global (A)ATSR data (1 km resolution), together
with 5 years of regional AVNIR-2 data (10 m resolution) to permit
on-the-fly Land Use / Land Cover Change analysis.

\subsection{Discretization}
As mentioned above, clustering issues include the possibility to consider different
distance measures among n-dimensional multi-spectral vectors
\citep{cerra2011compression}, ranging
from euclidean \citep{keogh2004towards} to spectral angle
\citep{kruse1993spectral, kruse1999fifteen}, from spectral correlation
\citep{decarvalo2000spectral} to information divergence measures
\citep{du2004new}.


KIM \citep{__datcu2003}, for instance, uses K-Means clustered
multi-spectral feature
histograms based on either euclidean or spectral angle distance
measures.
GeoIRIS \citep{shyu2007} exploits
histograms for panchromatic, grayscale RGB, and
near-infrared data.
\citet{maheshwary2009} apply K-Means on color moments and
co-occurrence matrices, then computes Euclidean distances and
validates the approach on three LISS III + multi-spectral satellite
images with $23.5~m$ resolution.

\marginparo{Hybrid hierarchical approaches}
Hybrid hierarchical approaches are also represented.
\citet{aksoy2006b} describes a hybrid hierarchical approach for
image content modeling and retrieval. First, scenes are decomposed
into regions using pixel-based classifiers and an iterative split-
and-merge algorithm. Next, spatial relationships of regions are
computed using boundary, distance and orientation information based on
different region representations. Finally, scenes are modeled using
attributed relational graphs that combine region class information and
spatial arrangements.

\marginparo{Particle Swarm optimization clustering}
Still working on limited amounts of data, a novel approach is represented in
\citet{DongX08} and \citet{Bedawi:2011:MCS:2027072.2027108},
Particle Swarm Optimization is used to 
classify remote sensing data:
classification rules are generated through
simulating the behaviors of bird flocking. Optimized intervals of each
band are found by particles in multi-dimension space, linked with land
use types for forming classification rules. Its performance with
respect to generalization issues still needs to be assessed in the
case of large archives.

\marginparo{Hierarchical partitioning}
Hierarchical partitioning is considered by a number of authors.
\citet{bahroun2010} propose a hierarchical visual thesaurus of the
regions provided by a region labeling criterion based on point pattern
analysis into homogeneous and textured regions for boosting the object
recognition. The labeling criterion is based on the spatial dispersion
of interest points in the region validated on a satellite image
database.

\subsection{Feature selection and ordering}
Techniques such as ICA \citep{comon_independent_1996} 
and PCA \citep{abdi2010principal}
are routinely used to generate more
efficient feature sets and to try and overcome the so called ``curse
of dimensionality''.

\marginparo{Feature selection and ranking}
In \citet{durbha2007image}, the relevance of a set of primitive features for
a particular land cover class or a combination of classes is then
assessed based on a wrapper-based genetic algorithm approach.
Furthermore, an induction algorithm is used along to arrive at an
optimal set of features.

\marginparo{Fisher LDA}
In \citet{aksoy2006}, spectral and textural features are quantized and
are used to train Bayesian classifiers with discrete non-parametric
density models. An iterative split-and-merge algorithm is used to
convert the pixel level classification maps into contiguous
regions. Then, the resulting regions are modeled using the statistical
summaries of their spectral, textural and shape properties. To
simplify computations and to avoid the curse of dimensionality during
the analysis of hyper-spectral data, \citet{aksoy2006} applies Fisher’s
linear discriminant analysis that finds a projection to a new
set of bases that best separate the data in a least-squares sense. It
applies principal components analysis that finds a projection to
a new set of bases that best represent the data in a least-squares
sense. The top 10 principal components are kept instead of the large
number of hyper-spectral bands.
Finally, discrete variables and a non-parametric model in the Bayesian
framework are considered where continuous features are converted to
discrete attribute values using the unsupervised K-Means clustering
algorithm for vector quantization. The number of clusters
(quantization levels) is empirically chosen for each feature.

\subsectiono{Compression--based measures}
Compression--based measures represent a niche with valuable
connections to the information--theoretical aspects of CBIR.
\marginparo{Normalized Compression Distance}
\citet{roman2011} use the Normalized Compression Distance as a measure
of similarity between two data files using the compression factor as
an approximation to the Kolmogorov complexity
\citet{cerra2011compression} instead proposes a Fast
Compression Distance that can be efficiently computed between images
and labels.

\section{Discussion}
\label{sec: discussion}
\begin{table}
\centering
\caption{Cited references per EO mining system function.}
\label{tab: decomposition_of_citations}
\footnotesize
\begin{tabular}{l}
\bf{Systems and architectures}\\ \hline
\citet{king2007}, \citet{guo2003}, \citet{tusk2003}, \citet{marchisio1998}, \citet{shyu2006}, \citet{tobin2006}, \\
 \citet{schroeder_interactive_2000}, \citet{datcu_knowledge_2002},
 \citet{__datcu2003}, \citet{molinier2006},\\
 \citet{tobin2006}, \citet{marchisio1998}, \citet{__datcu2003}\\ \\
\bf{Query specification}  \\ \hline
\citet{schroeder_interactive_2000}, \citet{__datcu2003},
\citet{shyu2007}, \citet{klaric2006}, \citet{barb2007}, \\
\citet{Ben-Yitzhak:2008:BBF:}, \citet{marchisio1998},
\citet{marchisio1999}, \citet{marchisio2000}\\ 
\citet{agouris1999environment}\\ \\
\bf{Labeling}   \\ \hline
\citet{__datcu2003}, \citet{shyu2007}, \citet{aksoy2006}, \citet{lienou2009}, \citet{lienou2009b},\\
  \citet{lienou2010}, \citet{durbha2004}, \citet{li2007b}, \citet{costache2007}, \citet{costache2008},\\
  \citet{Costache_Lienou_Datcu_2006}, \citet{ferecatu2005}, \citet{ferecatu2005b},
  \citet{ferecatu2007}, \citet{bordes2008},\\
 \citet{aksoy2009}, \citet{aksoy2010}, \citet{koperski2000},
  \citet{koperski2002}, \citet{blanchart_2011},\\
 \citet{alber2001}, \citet{costache2006}, \citet{barb2010},
  \citet{barb2010b}, \citet{ferecatu2007}, \\
  \citet{tuia2009active}, \citet{rital2008}\\ \\
\bf{Visualization}  \\ \hline
\citet{bratasanu2012}, \citet{molinier2006}, \citet{molinier2007}, \citet{chen2008}, \citet{guo2003},\\
  \citet{guo2006}, \citet{guo2006b} \citet{rivest2005solap} \\ \\
\bf{Query processing}  \\ \hline
\citet{durbha2008}, \citet{li2007geomatics}, \citet{vellaikal1995a}, \citet{__datcu2003},
  \citet{aksoy2004}, \\ 
\citet{durbha2005}, \citet{durbha2005c}, \citet{durbha2005b}, \citet{durbha2007image},
  \citet{durbha2009}, \\\citet{ruan2006}, \citet{datta2006}\\ \\
\bf{Indexing}  \\ \hline
\citet{__datcu2003}, \citet{shyu2006}, \citet{xiao2011},
\citet{tobin2006}, \citet{ayra1994}, \\
\citet{tobin2002},  \citet{shyu2007}, \citet{scott2011}\\ \\
\bf{Segmentation}  \\ \hline
\citet{guo2009}, \citet{gueguen2011}, \citet{durbha2004}, \citet{__datcu2003}, \citet{lienou2010},\\
  \citet{durbha2007image}, \citet{trias2008using}\\ \\
\bf{Primitive features} \\ \hline
\citet{__datcu2003}, \citet{aksoy2010}, \citet{guo2009}, \citet{molinier2006}, \citet{molinier2007},\\
  \citet{cerra2011compression}, \citet{shyu2007}, \citet{schroeder_interactive_2000},
 \citet{ aksoy2006}, \citet{tobin2006}, \\
\citet{pietikainen2000}, \citet{yao2003}   \citet{shah2005},
\citet{shah2007}, \citet{shah2006},\\
 \citet{li2007}, \citet{dellacqua2001}, \citet{silva2005}, \citet{walessa2000model},
  \citet{aiyesha2005}, \\
\citet{tusk2003}, \citet{koperski2002b}, \citet{aksoy2002}, \citet{aksoy2005},
  \citet{kalaycilar2008}, \\
\citet{pesaresi2001}, \citet{zhang2001}, \citet{matsakis1998},
\citet{molinier2005}\\ \\
\bf{Multi-sensor}  \\ \hline
\citet{molch2010}, \citet{molinier2006}, \citet{molinier2007},
\citet{molinier2008}\\ \\
\bf{Multi-temporal}  \\ \hline
\citet{heas2005modeling}, \citet{gueguen2007}, \citet{gueguen2008}, \citet{mantovani2009},
  \citet{natali2011},\\ \citet{molinier2007}, \citet{julea2006}, \citet{walter2004object}\\ \\
\bf{Discretization} \\ \hline
\citet{cerra2011compression}, \citet{__datcu2003}, \citet{shyu2007}, \citet{maheshwary2009},
  \citet{aksoy2006b}, \\
\citet{DongX08}, \citet{Bedawi:2011:MCS:2027072.2027108}, \citet{durbha2007image},
  \citet{aksoy2006}, \citet{roman2011}, \\
 \citet{mountrakis2011support}
\end{tabular}
\end{table}

\marginparo{A lively domain}
EO mining evidently is a very lively research domain. Yet the issues
to be overcome are still significant with respect to both basic
research and operations. 
With particular reference to the main systems represented in the
literature and whose attributes are reported in 
table~\ref{tab: eo_mining_system_decomposition}, 
the issues of domain--specific (e.g. agriculture,
disaster, urban etc) performance connected to basic architectures need
to be taken into account together with aspects such as
interoperability and scalability.
\marginparo{Domain--specific performances}
The issue of domain--specific performances involves the specific
optimizations that have been opted for in the engineering trade--offs
typical of system design. Each one of the considered systems can also
be seen as determined in its strenghts and limitations by the choices
taken with the aim of a specific application domain.

\marginparo{Possibilities vs ease of use: GeoBrowse}
The GeoBrowse system allows
experimentation in a high--level statistically oriented environment
with rich data plotting and explorative data analysis capabilities.
Examples in the published literature, though, mainly refer to LANDSAT
data and to simple land cover classes. 
The relative richness of the operating environment might have proved a
limiting factor as well as a competitive advantage for the system.
Here the dilemma at the base of the design seems to have been
flexibility versus ease of use.

\marginparo{Performance vs flexibility: GeoIRIS}
GeoIRIS is clearly optimized for metric resolution images of man--made
objects. Security appears to be the central considered
application. This well--specified focus has visibly impacted the
system design, for instance in the way very effective yet perhaps
scarcely flexible
indices are built into the DBMS that serves as a back--end for the
system.
In the case of this system, performance and flexibility seem to have
been the two competing considerations.

\marginparo{: $I^3KR$}
A peculiarity of the $I^3KR$ system is its focus on the modeling of 
information sources by hybrid domain-specific
ontologies used for data exploration and integration
tasks.
The tool has been exercised in the context of rapid mapping
for disaster management in coastal areas: ontologies for Landsat and
MODIS imagery based on the Anderson classification system
have been developed. Further ontologies for land cover characteristics
have been
conceptualized in the IGBP ontology and concepts in the hydrology
domain have been formalized.
A limitation in this approach might be related to the way 
the flexibility of the system is only
obtainable by the proper modelling of the domain ontologies through
experimentation by advanced users.

\marginparo{Flexibility vs simplicity: KEO/KIM}
KIM/KEO is probably the most flexible and interoperable among the
considered systems.
For instance, it allows users to easily include new feature
extraction tools in the system. 
It has been applied in scenarios ranging from large archive management
to flood mapping.
Yet, focus on a specific application domain seems to be lacking.
This in turn has made it probably difficult to focus on performance.
KIM/KEO seems to be the result of choices that are the opposites of
those at the basis of GeoIRIS. In this case, flexibility has been
chosen on simplicity and sheer performance.

\marginparo{Performance vs flexibility: RBIR}
The RBIR system is based on an innovative agent--based architecture.
It is demonstrated on a large set of cover types ranging
from forest to industrial. The primitive feature set considered is
efficient in the considered application domain, yet the authors
describe the need to extend it in order to support multi--temporal
queries and to improve specificity via color information.
Again, the performance might have been obtained at the partial expense of
flexibility.

\marginparo{Experimentation only: PicSOM}
The key application for the PicSOM system is related to the detection
of buildings and the monitoring of construction and destruction events
in couples of images. Experimentations on archives with larger data volumes
and a broader spectrum of applications will be needed to evaluate the
scalability of its performance to large scale operations conditions.

\marginparo{Scalability}
The issue of scalability is of course a key one for all of the
considered systems. KIM/KEO and GeoIRIS are arguably the tools that
have been experimented with on the largest scales. Yet, the
quantitative measurement of retrieval performance in the case of very
large datasets requires the availability of ground truth information
that can be extremely expensive to acquire. New strategies need
to be devised in this regard \citep{quartulli_2012_esa_eusc}.

\marginparo{Only mining well-established archives?}
The picture that emerges shows that most of the available
tools seem to be focusing on well--established archives from past
missions rather than on currently active ones. 
\marginparo{Issues: missing strange sensors}
The attributes in table~\ref{tab: eo_mining_system_decomposition} show
how most of the 
tools deal with decametric data from either electro-optical
multi-spectral sensors.
Limited support is available for more specialized sensor classes such
as SAR and LIDAR.
Multi-temporal analysis is present only in a very small handful of
systems.

\marginparo{Adapting to Sentinels}
While this is perfectly understandable based on historical reasons, 
it nevertheless might
represent a significant limiting factor in the overall efficacy of
novel missions with their increased data rates.
It is very important that systems
that are operational on Quickbird or SPOT data are effectively
adapted to be used e.g. on the upcoming Sentinel data.

\marginparo{Issues: missing high resolution}
The provision of systems such as those in~\cite{tobin2006,
  costache2007} to understand the metric resolution data from
sensors such as the WorldView, QuickBird, TerraSAR-X, COSMO Skymed,
Sentinel systems is urgently needed, given their present and expected
data production rates.

\marginparo{Issues: missing multi-sensor, multi-temporal}
Furthermore, for objectives related to rapid mapping and large-scale
scene understanding in connection with large product archives, more
effective tools
for multi-sensor and multi-temporal EO mining and analysis are required.

\marginparo{Usage patterns}
The specific usage patterns related to the management and exploitation
of very large EO archives are to be taken into consideration as
essential to the fulfillment of the promise of Earth observation for
the next decades, 
\marginparo{eobench}
and the issue of quantitatively evaluating the different approaches in
order to measure their relative merits needs to be considered.




\marginparo{Issues: missing noise models}
A significant limitation of a number of approaches (e.g. the one in
\citet{mantovani2009} and \citet{natali2011} with respect to the one in
\citet{julea2006}) is the lack of prevision for incomplete knowledge of
the data: the approaches are purely deterministic and does not foresee
sound procedures for sensor noise and error treatment.

\marginparo{The first diagram}
Going back to the plot in figure \ref{fig: cited_per_year}, we
note that the temporal distribution of the considered citations shows
that after a seminal phase that extends to the early 1990s, and after
an initial implementation in 2000-2005, the second half of the last
decade has been characterized by an explosion of activity.
This growth seems now to be receding in a phase in which
archive owners and operators are transitioning to operational
systems for automatic data annotation connected directly to satellite
payload ground segments.

\section{Conclusions}
We have analyzed the state of the art of content-based
retrieval in Earth observation remote sensing image archives
with specific attention to complete systems showing promise
for operational implementation.

A generic system model has been presented and the functionality
  decomposition it defines has been used as a basis for the analysis of
  published approaches. 

The different paradigms at the basis of the main system
families have been introduced.
The approaches taken have been analyzed, focusing in particular on the
phases after primitive feature extraction. 
The solutions envisaged for the issues related to feature
simplification and synthesis, indexing, semantic labeling have been
reviewed.
The methodologies for query specification and execution have been
analyzed.

A large variety of approaches is present in the
literature. Unfortunately, no general benchmarking tool and dataset
are available for their evaluation.
Quantitative benchmarking methods are needed to address the question
of evaluating the efficacy of the engineering choices they embody.

\end{document}